\documentclass[10pt,journal]{IEEEtran}

\usepackage{float}
\usepackage{xcolor,colortbl} 
\usepackage{amsmath,amsfonts} %
\usepackage{graphicx}%
\usepackage{textcomp}%
\usepackage{xcolor}%

\usepackage{array}%
\usepackage{pifont}%

\usepackage{amsthm}%
\usepackage{bm}

\usepackage[linesnumbered,ruled,vlined]{algorithm2e}%
\usepackage{setspace}%

\usepackage{multirow}%
\usepackage{siunitx}
\usepackage{footnote}%
\usepackage[utf8]{inputenc}
\usepackage[english]{babel}

\usepackage{blindtext}

\usepackage{dblfloatfix}%

\usepackage{xcolor,colortbl} 
\usepackage{ctable} 
\graphicspath{ {image/} }

\usepackage{pifont}
\newcommand{\xmark}{\ding{55}}

\usepackage{booktabs}
\usepackage{makecell}
\usepackage{comment}

\usepackage[normalem]{ulem}
\usepackage{soul}

\def\BibTeX{{\rm B\kern-.05em{\sc i\kern-.025em b}\kern-.08em
    T\kern-.1667em\lower.7ex\hbox{E}\kern-.125emX}}
\usepackage{balance}

\begin{document}


\title{SUBARU: A Practical Approach to Power Saving in Hearables Using \textbf{SUB}-Nyquist \textbf{A}udio \textbf{R}esolution \textbf{U}psampling}

\author{
Tarikul Islam Tamiti, Sajid Fardin Dipto, Luke Benjamin Baja-Ricketts, David C Vergano, Anomadarshi Barua$^1$, Department of Cyber Security Engineering, George Mason University, USA.
}


\maketitle

\begin{abstract}
Hearables are wearable computers that are worn on the ear. Bone conduction microphones (BCMs) are used with air conduction microphones (ACMs) in hearables as a supporting modality for multimodal speech enhancement (SE) in noisy conditions. However, existing works don't consider the following practical aspects for low-power implementations on hearables: (i) They do not explore how lowering the sampling frequencies and bit resolutions in analog-to-digital converters (ADCs) of hearables jointly impact low-power processing and multimodal SE in terms of speech quality and intelligibility. And (iii) They don't process signals from ACMs/BCMs at a sub-Nyquist sampling rate because, in their frameworks, they lack a wideband reconstruction methodology from their narrowband parts. We propose SUBARU (\textbf{Sub}-Nyquist \textbf{A}udio \textbf{R}esolution \textbf{U}psampling), which achieves the following: SUBARU (i) intentionally uses sub-Nyquist sampling and low bit resolution in ADCs, achieving a 3.31x reduction in power consumption;  and (ii) achieves streaming operations on mobile platforms and SE in in-the-wild noisy conditions with an inference time of 1.74ms and a memory footprint of less than 13.77MB. 
\end{abstract}


\vspace{-1em}
\section{Introduction}
A hearable device is a wearable computer that is worn on the ear. 
Initially popularized through earbuds and headphones, the hearables market has expanded into diverse applications, including health monitoring, augmented reality, and voice assistance, with a projected size of \$90 billion by 2030 \cite{hearables_market_report_2024}. 

\begin{figure}[t]
\vspace{-0.5em}
    \centering
    \includegraphics[width=0.49\textwidth,height=0.22\textheight]{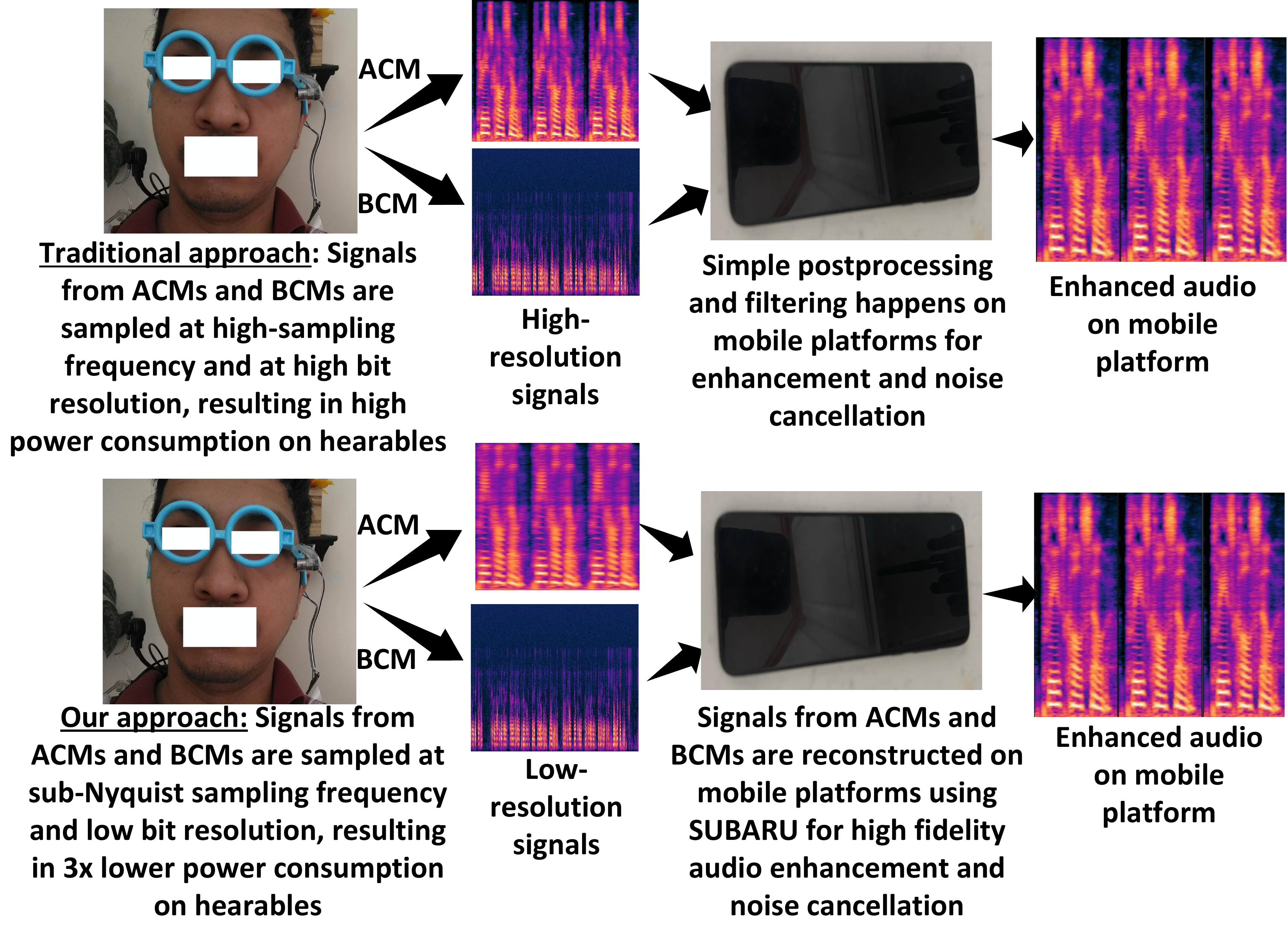}
    \vspace{-01.585em}
    \caption{We propose to use sub-Nyquist sampling and low bit resolution on hearables in split architecture where audio will be reconstructed with low latency and high fidelity on mobile platforms.}
    \label{fig:overview}
    \vspace{-01.25em}
\end{figure}

Traditionally, air conduction microphones (ACMs) are used in hearables that can easily pick up background noise, degrading speech quality. 
Recently, \textit{bone conduction microphones (BCMs) and accelerometers} are proposed with ACMs as conditional signal enhancer for multimodal speech enhancement (SE) in noisy conditions \cite{sui2024tramba, tagliasacchi2020seanet, wang2022end}. 
However, any SE algorithms do not consider the following practical aspects of low-power and low-memory applications in hearables:



\begin{itemize}

\item \textbf{Lowering sampling frequency and bit resolution:} Analog audio or vibration signals from ACMs and BCMs, respectively, are first sampled and digitized at Nyquist rates (greater than 16 kHz) and over 12-bit resolutions by the analog-to-digital converter (ADC). After sampling, audio codecs compress data to reduce the bitrate, saving transmission energy and bandwidth. Later, multimodal SE algorithms are applied on the decompressed data on connected mobile platforms (i.e., cell phones, see Fig. \ref{fig:overview}). \textit{However, they do not explore how lowering the sampling frequency and bit resolutions in ADCs of hearables jointly impact low-power processing and multimodal SE.}

\item \textbf{Model complexity:} State-of-the-art (SOTA) multimodal SE algorithms use generative adversarial networks (GANs) over U-Nets because of GANs' promising performance over U-Nets. \textit{However, GANs often exhibit high complexity with numerous associated parameters (on the order of hundreds of millions), thereby imposing significant constraints on efficiency in resource-constrained mobile platforms connected to hearables.}



\end{itemize}

In this paper, we aim to enable multimodal SE that adapts effectively to the unique low-power features by \textit{jointly reducing the sampling frequency to the sub-Nyquist range with lower bit resolutions in hearables}. We name our methodology SUBARU (\textbf{Sub}-Nyquist \textbf{A}udio \textbf{R}esolution \textbf{U}psampling). 
SUBARU jointly leverages bandwidth extension (BWE) with multimodal SE  to recover missing high-frequency parts from the sub-Nyquist sampled and lower bit-resolution audio in noisy conditions, restoring audio quality and intelligibility (\textbf{see Fig. \ref{fig:overview}}). 
SUBARU has the following four key design elements:

\textbf{First},  SUBARU is designed to be deployed in a \textit{split architecture}, where \textit{hearables only} run sub-Nyquist sampling in low bit resolution on both ACMs and BCMs, and \textit{mobile platforms connected to hearables only} run the joint BWE and multimodal SE algorithms to reconstruct the high-resolution audio. The low latency of SUBARU while doing joint BWE and multimodal SE enables streaming enhancement that makes SUBARU deployable on mobile platforms.

\textbf{Second}, GANs can reconstruct better perceptual audio through adversarial loss functions, and U-Nets generally have a lower memory footprint and faster convergence while training. To get the best out of these two domains, SUBARU adopts multi-scale and multi-period loss functions in its U-Net architecture. 
\textit{SUBARU achieves similar perceptual quality to GANs with smaller parameters (saving tens of millions) and stable training, enabling efficient generation in mobile platforms.}

\textbf{Third}, 
Waveform-based methods are performed at the sample-point level, leading to a simpler architecture but relatively lower generation efficiency when compared to spectrum-based methods with frame-level operations. \textit{To get the benefits from both methods, SUBARU adopts a hybrid architecture by merging both waveform-based and spectrum-based methods, enabling joint training in both spectrum and raw waveform.} 

\textbf{Fourth}, As recent research \cite{yin2020phasen, zhang2025wearse} highlights the importance of phase reconstruction for improved perceptual quality, \textit{the joint training feature of SUBARU in both spectrum and waveform domains enables the use of instantaneous phase and group delay anti-wrapping losses to reconstruct clean phases in noisy conditions.} This also helps to provide GAN-like perceptual quality in U-Net-based SUBARU, keeping the model size small and efficient. 

This paper implements SUBARU focusing on the above four design elements, and extensive experiments are conducted to evaluate its performance in real-time settings. The summary of the evaluations is written below:

\begin{itemize}


\item SUBARU has been extensively evaluated against ten baselines. Five of them are GAN-based (i.e., SEANet, AERO, HiFi++, EBEN, and NVSR), one is diffusion-based (i.e., NU-Wave), and four are U-Net-based models (i.e., TFiLM, AFiLM, ATS-UNet, and VibVoice). SUBARU is evaluated using six evaluation metrics: LSD, VISQOL, NISQA-MOS, SI-SDR, PESQ, and STOI. The full form of the abbreviations is given in Section \ref{subsec:Evaluation Metrics}.

\item \textit{By reducing the sampling frequency and bit resolution from \{24 kHz, 12-bit\} to \{4 kHz, 8-bit\}, SUBARU achieves a 3.31x reduction in power consumption, ideally increasing the battery life by $\sim$3.31x for hearables.}

\item SUBARU is evaluated on both speech and music datasets and is superior to the SOTA U-Net models.  Moreover, SUBARU achieves better performance than SOTA GANs under noisy conditions with the lowest inference time (i.e., 1.74 ms on a desktop with GPUs). Specifically, SUBARU needs 3.61x less inference time compared to HiFi++ and AERO (the two best-performing models). 
Thanks to the multi-scale and multi-period loss functions.


\item SUBARU has an inference time of 71 ms on a mobile platform (i.e., Google Pixel7), which is smaller than the 150 ms threshold set by the International Telecommunication Union \cite{ITU-G114}, indicating its capability of streaming operation from hearables to mobile platforms. The above results are further verified by using Samsung Galaxy S21.


\end{itemize}

\vspace{-0.35em}
\section{RELATED WORK}
\label{sec:related_work}
\vspace{-0.1em}

As SUBARU can run both single-modal (i.e., ACM only) and multimodal (i.e., ACM with accelerometer or BCMs) SE with variable noise, we divide our discussion into two sections: (i) Bandwidth extension, and (iii) Multimodal SE.

\vspace{-0.8em}
\subsection{Low-power and low-memory BWE on hearables}
\label{subsec:Bandwidth extension in hearables}

BWE is an ideal fit for audio reconstruction from sub-Nyquist sampling. Recently, neural networks become the SOTA solutions for BWE, ranging from pure feedforward networks \cite{li2015deep, kuleshov2017audio, gupta2019speech, kuleshov2017audio, li2021real, su2021bandwidth, han2022nu, lagrange2020bandwidth, li2018speech, kumar2020nu, eskimez2019speech,li2015deep} to generative solutions \cite{moliner2022behm, moliner2023solving, liu2024audiosr, moliner2024blind, moliner2024diffusion}. However, due to the large memory and power requirements, most of them are not directly suitable for hearables. To solve this problem, Li et al. proposed ATS-UNet \cite{li2022enabling}, which is built upon the TFiLM \cite{birnbaum2019temporal} backbone, resulting in a smaller footprint by compromising the audio quality. To improve audio quality, the model size needs to be increased. Therefore, TRAMBA (IMWUT, 2025) \cite{sui2024tramba} deploys BWE models not on hearables but on mobile platforms, as mobile platforms have higher hardware capabilities than hearables. TRAMBA is based on U-Net and achieves a smaller size without sacrificing performance. However, TRAMBA has the following limitations:

\begin{itemize}
\item  It operates on raw waveform and performs at the sample-point level, leading to relatively lower generation efficiency when compared to spectrum-based methods \cite{mandel2023aero}.
\item  It does not consider clean phase reconstruction in noisy conditions while reconstructing high-resolution audio from a low-resolution signal. 
\item \textit{Last and most importantly,} TRAMBA and no other SOTA work consider the effect of different bit resolutions and quantization noise during sub-Nyquist sampling in their BWE algorithms (see Section \ref{subsec:Audio quality at different sub-Nyquist sampling} for details).

\end{itemize}


While recent GANs (i.e., SEANet \cite{tagliasacchi2020seanet}, AERO \cite{mandel2023aero}, EBEN \cite{hauret2023eben}, NVSR \cite{liu2022neural}) and diffusion-based BWE methods (i.e., \cite{ho2020denoising}, NU-Wave \cite{lee2021nu}, NU-Wave 2 \cite{han2022nu}) have demonstrated promising performance, they still require numerous time steps, a.k.a. latency, in the reverse process for waveform reconstruction (on the order of hundreds of milliseconds) and often exhibit high complexity with numerous associated parameters (on the order of hundreds of millions), thereby imposing significant constraints on generation efficiency in low-power and low-memory resource-constrained systems like hearables. Moreover, they have not been properly explored for joint BWE and multimodal SE in hearables for noisy conditions. 


\vspace{-1em}
\subsection{Joint BWE and multimodal SE on hearables}
\label{subsec:Multi-modal speech enhancement in hearables}

BCMs are commonly used as an accessorial enhancer for ACMs. BCMs can typically consist of two sensors: vibration sensors and accelerometers. 
SEANet \cite{tagliasacchi2020seanet}, a GAN model, uses accelerometers only for multimodal SE, without resorting to explicit joint BWE with SE. 
HiFi++ \cite{kim2023hifi++}, a GAN framework, is proposed for joint BWE and SE of ACMs, but has not been adapted to the multimodal domain (i.e., works with audio signal only).  ClearSpeech \cite{ma2023clearspeech} leverages the different acoustic properties captured by in-ear and out-ear microphones to enhance ACM's signals, but does not consider BWE, hence, it does not work at the sub-Nyquist sampling rate. EarSpeech \cite{han2024earspeech}  enhances airborne speech by analyzing the cross-channel correlation between in-ear and airborne signals. VibVoice \cite{he2023vibvoice}  employs bone-conducted vibrations, showing potential for environments with heavy background noise. 

All the above models have either one or multiple of the following limitations for which they are not suitable for our low-power hearables: \textbf{(i)} The joint BWE and SE models are typically evaluated for single modality (i.e., audio only) and have not been extended to multimodality. \textbf{(ii)} The multimodal SE models use narrowband signals from BCMs (i.e., sampling frequency 4 kHz or greater) and full wideband (i.e., sampling frequency 16 kHz or greater) audio signals from ACMs. Therefore, they cannot handle \textit{audio signals} at a sub-Nyquist sampling rate because of the absence of the BWE algorithm in their frameworks. 
And, \textbf{(iii}) These methods are not designed to be deployed on split architecture, where hearables run only sub-Nyquist sampling in both ACMs and BCMs, and mobile platforms run the BWE and multimodal SE algorithms. 
A summary of limitations is shown in Table \ref{table:summaryjointBWE}. We address all these limitations in our proposed SUBARU (see Section \ref{sec:performance evaluation}).

\begin{table}[ht!]
\scriptsize
    \centering
    \vspace{-01.1851800em}
    \caption{A summary of limitations.}
    \vspace{-0.81800em}
    \setlength{\tabcolsep}{1.6pt}
    \begin{tabular}{l | l|l|l|l}
    \hline
        \cellcolor [gray]{0.85}\textbf{Model name} & \cellcolor [gray]{0.85}\textbf{Architecture} & \cellcolor [gray]{0.85}\textbf{BWE} & \cellcolor [gray]{0.85}\textbf{Single-modal SE} & \cellcolor [gray]{0.85}\textbf{Multimodal SE}\\ 
        \hline
        \hline
        ATS-UNet \cite{li2022enabling} & U-Net &\checkmark &  \checkmark & \xmark  \\ 
        \hline
        TFiLM \cite{birnbaum2019temporal} & U-Net & \checkmark & \checkmark & \xmark \\
       \hline
        AFiLM \cite{rakotonirina2021self} & U-Net & \checkmark & \checkmark & \xmark \\
       \hline
        TRAMBA \cite{sui2024tramba} & U-Net & \checkmark & \checkmark  & \xmark \\ 
        \hline
         ClearSpeech \cite{ma2023clearspeech} &  U-Net & \xmark &  \xmark &  \checkmark  \\ 
        \hline
        VibVoice \cite{he2023vibvoice} &  U-Net & \xmark &  \xmark &  \checkmark  \\  \hline
        \hline
         SEANet \cite{tagliasacchi2020seanet} & GAN & \xmark & \checkmark  &  \checkmark  \\ 
         \hline
         AERO \cite{mandel2023aero} & GAN & \checkmark  &  \checkmark &  \xmark  \\ 
        \hline
        EBEN \cite{hauret2023eben} & GAN & \checkmark  &  \checkmark &  \xmark \\ 
        \hline
         HiFi++ \cite{kim2023hifi++} &  GAN & \checkmark  &  \checkmark &  \xmark  \\ 
        \hline
        NVSR \cite{liu2022neural} & U-NET + GAN & \checkmark  & \checkmark  &  \xmark  \\ 
        \hline
        NU-Wave \cite{lee2021nu} & Diffusion & \checkmark  &  \checkmark &  \xmark  \\ 
        \hline
         \hline
        \textbf{Proposed SUBARU}  &  \textbf{U-Net} & \textbf{\checkmark} &  \textbf{\checkmark} &  \textbf{\checkmark}  \\ 
        \hline
    \end{tabular}
    \vspace{-0.8520em}
    \label{table:summaryjointBWE}
\end{table}

Please note that a 4-page version of this manuscript has been submitted to Interspeech 2026, which is currently under review and is attached as a supporting document with this manuscript. This manuscript has the following improvements:

\ul{\textbf{i)}} We add a new Time Enhancement Network (see Section \ref{subsec:time enhancement network}) and an additional Multi-resolution STFT loss (see Section \ref{subsec:Combined loss in the frequency, time, phase domains}) to improve the performance of our framework. We achieve better LSD (0.84 vs 0.87), VISQL (4.35 vs 4.15), NISQA-MOS (4.19 vs 4.13), SI-SDR (17.94 vs 16.99), with similar PESQ (3 vs 2.99) and STOI (0.90 vs 0.90).

\ul{\textbf{ii)}} We use ten different models as baselines for comparison in this manuscript, compared to only six baselines in the 4-page version (see Section \ref{subsec:Base models}).

\ul{\textbf{iii)}} We show a detailed comparison of how Mamba improves efficiency compared to Transformers in Sections \ref{subsec:Spectral enhancement network} \& \ref{subsec:time enhancement network}.

\ul{\textbf{iv)}} In this manuscript, as the collected data amount for {ACM, BCM} pair is not sufficient, we create a synthetic dataset for better training using six different SEANet models for three different bit resolutions of 12, 10, and 8 bits (see Section \ref{subsec:Dataset collection}). 

\ul{\textbf{v)}} We show a detailed breakdown of our model by size, parameters, and inference time in Section \ref{subsec:Breakdown of SUBARU size and parameters}.

\ul{\textbf{vi)}} This manuscript shows how SUBARU does streaming operation frame-by-frame in Section \ref{subsec:Streaming operation of SUBARU}.

\ul{\textbf{vii)}} This manuscript shows a detailed evaluation separately on speech enhancement and bandwidth extension (see Sections \ref{subsec:Evaluation for BWE} and \ref{subsec:Evaluation for joint BWE and multi-modal SE}), which are not present in the 4-page version.

\ul{\textbf{viii)}} This manuscript shows a detailed evaluation separately only on different sub-Nyquist sampling with music dataset (see Sections \ref{subsec:Evaluation for only BWE with different sub-Nyquist sampling}), which is not present in the 4-page version.

\ul{\textbf{ix)}} This manuscript shows a detailed evaluation on live noise data inside and outside of the lab in bus-ride, classroom, and car ride scenarios (see Section \ref{subsec:Evaluation on live noise data}), which is not present in the 4-page version.  

\vspace{-0.520em}
\section{PRELIMINARY}
\label{sec:preliminary}
\vspace{-0.10em}

\subsection{Power at sub-Nyquist  frequencies and  bit resolutions}
\label{subsec:Power at different sub-Nyquist sampling }

The power consumption of ADCs in hearables increases with higher sampling rates and resolutions, following $P = k \cdot f_s \cdot 2^N$, where $P$ is the consumed power, $k$ is a proportionality constant, $N$ is the bit resolution, and $f_s$ is the sampling frequency of ADCs. Traditional methods require ADCs to operate at high sampling frequencies (i.e., $>$16 kHz) and bit resolution (i.e., 12-24 bits) to accurately capture wideband audio in hearables. However, lowering sampling frequencies and bit resolution offers a powerful approach to reducing energy consumption in low-power applications like hearables. 

To support this claim, we conduct experiments with an ACM (i.e., part $\#$ B\&K Type 4192 \cite{bk4192_datasheet}) and an in-built ADC of NRF52840 \cite{nordic_nrf52840}. We get a relative power savings of 2.45x between \{16 kHz, 12-bits\} vs \{4 kHz, 8 bits\} computations. 

Please note that sub-Nyquist sampling and low-resolution bits in ADCs will reduce the audio quality in hearables. Therefore, our proposed SUBARU implements joint BWE and multimodal SE in mobile platforms (i.e., cell phones) to recover audio at high resolution. The increase in power consumption on mobile platforms to run additional BWE and SE algorithms is negligible, as SOTA mobile platforms typically have 100x or more battery capacity compared to hearables (i.e., Samsung Galaxy Buds2 Pro has a 50 mAh battery compared to a 5000 mAh battery in Samsung Galaxy S24 Ultra). An  explanation of power performance on both hearables and mobile platforms is discussed in Section \ref{subsec:Evaluation for power consumptio}.


\begin{figure}[h!]
\vspace{-0.65em}
    \centering
    \includegraphics[width=0.49\textwidth,height=0.15\textheight]{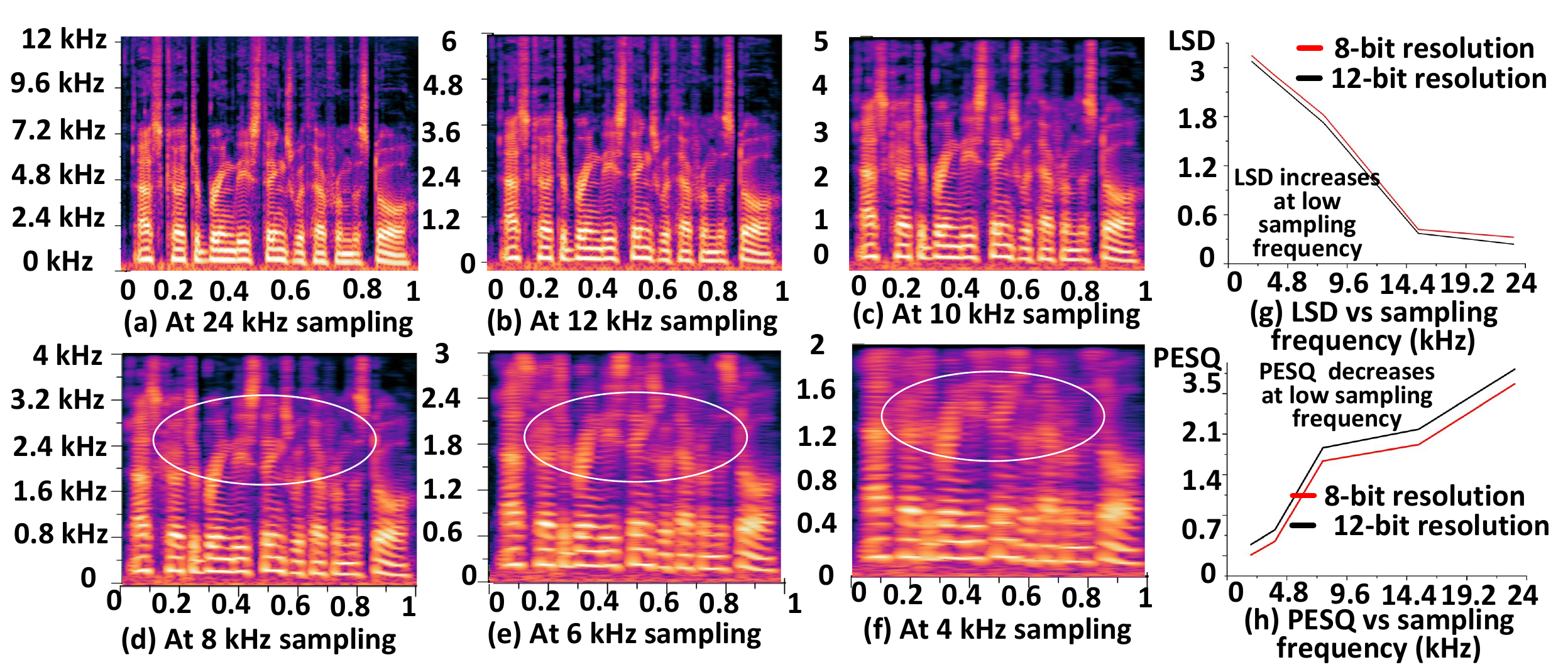}
    \vspace{-01.85em}
    \caption{A 48 kHz sampled reference signal is down sampled at (a) 24 kHz, (b) 12 kHz, (c) 10 kHz, (d) 8 kHz, (e) 6 kHz, and (f) 4 kHz. (g) The LSD increases and (h) PESQ decreases from higher to lower sampling frequencies. (g and h) The 8-bit audio has less quality than 12-bit audio in terms of PESQ and LSD. The circle marks indicate that downsampling reduces signal quality.}
    \label{fig:sub-Nyquist sampling frequencies and bit resolutions}
    \vspace{-0.95em}
\end{figure}

\begin{figure*}[h!]
\vspace{-0.63em}
    \centering
    \includegraphics[width=0.9\textwidth,height=0.15\textheight]{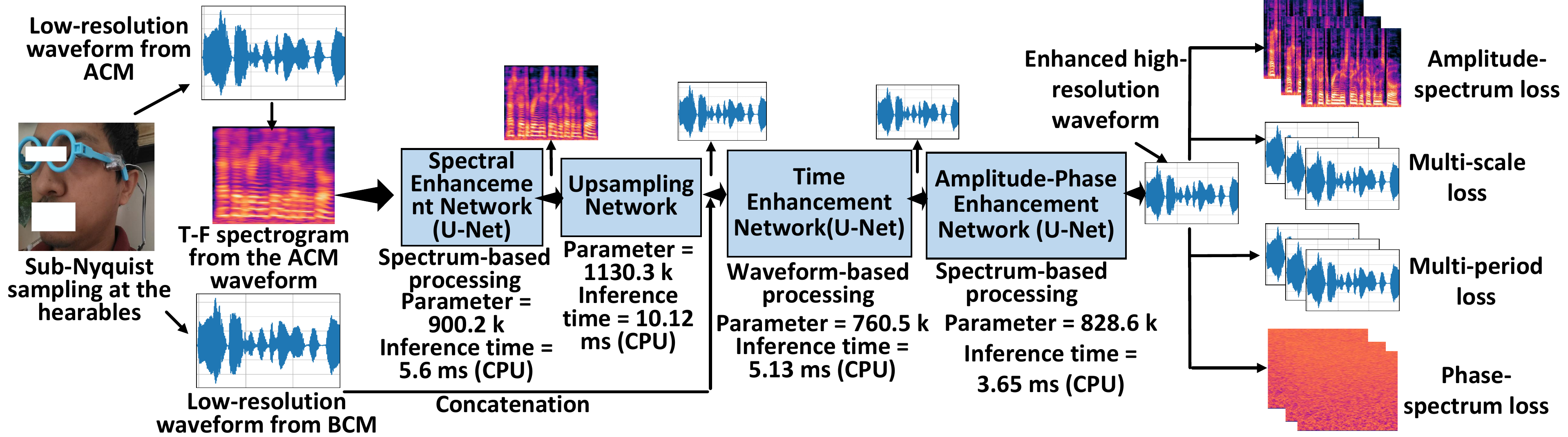}
    \vspace{-0.863em}
    \caption{Overview of the SUBARU architecture.}
    \vspace{-01.73em}
    \label{fig:basic_model}
\end{figure*}

\vspace{-0.75em}
\subsection{Audio at sub-Nyquist frequencies and bit resolutions}
\label{subsec:Audio quality at different sub-Nyquist sampling}

To gain an understanding of how the bit resolutions and sub-Nyquist sampling frequencies impact the audio quality in hearables, we conduct several pilot studies. Fig. \ref{fig:sub-Nyquist sampling frequencies and bit resolutions} presents time-frequency (T-F) spectrograms of 1s speech signals while a volunteer speaks wearing hearables at $N$ = \{12-bit, 8-bit\} resolutions for different sampling frequencies, $f_s$ = \{24 kHz, 12 kHz, 10 kHz, 8 kHz, 6 kHz, 4 kHz\}. Fig. \ref{fig:sub-Nyquist sampling frequencies and bit resolutions} indicates that with the decrease of the sampling frequency, the higher frequency components (a.k.a. formants)  of audio are deprecated, resulting in lower sound quality (i.e., lower Perceptual Evaluation of Speech Quality (PESQ) and higher Log Spectral Distance (LSD)). Moreover, the lower bit resolution increases the quantization noise, resulting in lower audio quality. Therefore, the lower bit resolution of $N$  = 8 bits provides lower quality audio compared to the 12-bit resolution. SUBARU considers the impact of the low bit resolution in BWE algorithms that is absent in SOTA research.

\vspace{-0.73em}
\section{SUBARU ARCHITECTURE DESIGN}
\label{sec:SUBARU ARCHITECTURE DESIGN}
\vspace{-0.2em}

SUBARU is engineered to achieve the following objectives: 

\begin{itemize}
\item SUBARU will handle dynamic interferences from extremely noisy conditions in mobile scenarios using joint BWE and multimodal SE algorithms.

\item SUBARU will process input audio signals in both the spectrum and waveform domains and enhance both the amplitude and phase of the time-frequency (T-F) spectrum to output high-quality speech signals from sub-Nyquist sampled data (i.e., low-resolution audio).

\item SUBARU will enable streaming enhancement and low-power and low-memory solutions that will make SUBARU deployable on mobile platforms. 

\end{itemize}


SUBARU adopts a U-Net-based model, which is explained below. Figure \ref{fig:basic_model} illustrates the general framework of SUBARU.

\vspace{-0.90em}
\subsection{Spectral enhancement network}
\label{subsec:Spectral enhancement network}

\begin{figure}[h!]
    \centering
    \includegraphics[width=0.49\textwidth,height=0.14\textheight]{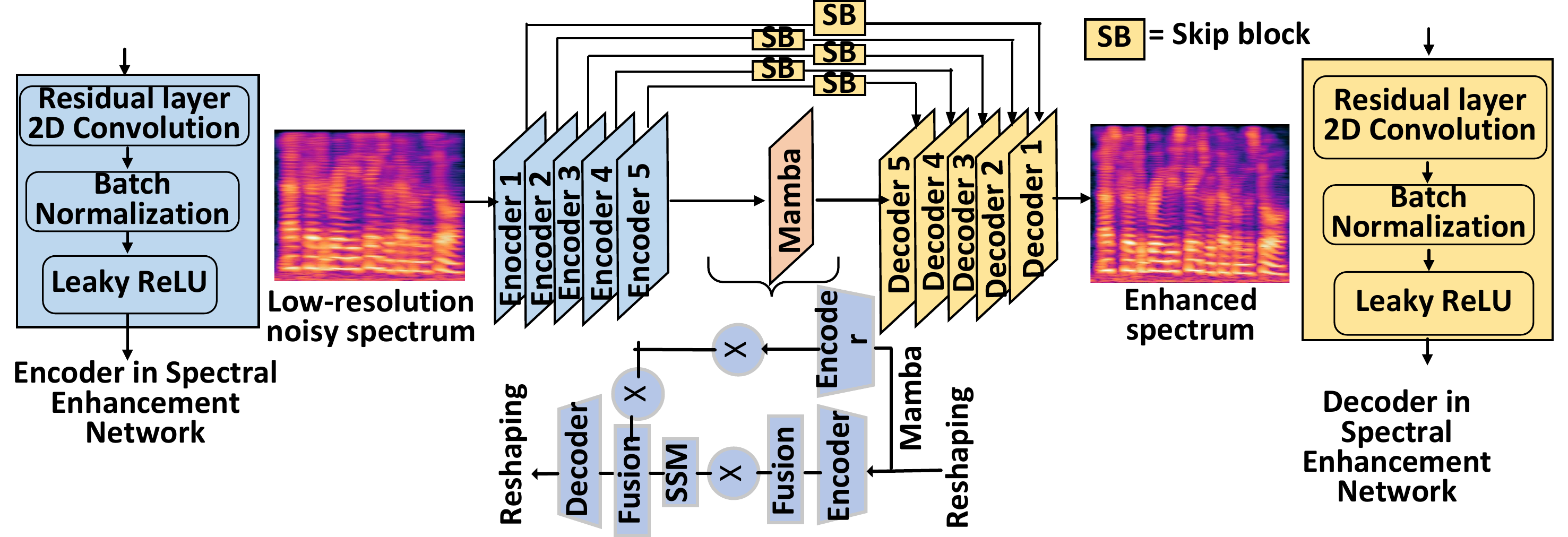}
    \vspace{-02em}
    \caption{SEN enhances noisy spectrograms for waveform-based processing.}
    \label{fig:spectral enhancement network}
\end{figure}

The spectral enhancement network (SEN) is the initial part of SUBARU that takes the spectrogram of the low-resolution and noisy audio from ACMs as inputs (see Fig. \ref{fig:basic_model} and \ref{fig:spectral enhancement network}). 
The network also works as the initial stage of converting the spectrogram into a waveform. 

The network architecture (see Fig. \ref{fig:spectral enhancement network}) consists of a 2D convolution framework in U-Net, featuring 5 residual layers as encoders (Enc), 5 residual layers as decoders (Dec), and a Mamba \cite{gu2023mamba} block as a bottleneck layer. The details of each layer are outlined in Table \ref{table:spectral enhancement network}. Each residual layer incorporates batch normalization followed by leaky ReLU activations. Dilations are used to increase the receptive fields that help to capture local features over a dilated window, resulting in capturing inter-phoneme dependencies in audio signals. 

\begin{table}[ht!]
\vspace{-01.21800em}
\scriptsize
\setlength{\tabcolsep}{0.8pt}
    \centering
    \caption{The details of the spectral enhancement network.}
    \vspace{-0.51800em}
    \begin{tabular}{l | l|l|l|l | l|l|l|l | l|l|l}
    \hline
        \cellcolor [gray]{0.85}\textbf{Layer} & \cellcolor [gray]{0.85}\textbf{Enc1} & \cellcolor [gray]{0.85}\textbf{Enc2} & \cellcolor [gray]{0.85}\textbf{Enc3} & \cellcolor [gray]{0.85}\textbf{Enc4} & \cellcolor [gray]{0.85}\textbf{Enc5} & \cellcolor [gray]{0.85}\textbf{Dec1} & \cellcolor [gray]{0.85}\textbf{Dec2} & \cellcolor [gray]{0.85}\textbf{Dec3} & \cellcolor [gray]{0.85}\textbf{Dec4} & \cellcolor [gray]{0.85}\textbf{Dec5} &  \cellcolor [gray]{0.85}\textbf{Bottleneck}\\ 
        \hline
        \hline
        Kernels & 8 &  16 & 24 & 32 & 64 & 64 & 32 & 24 & 16 & 8 &  \\ 
        \hline
       Kernel size & 4x4 & 4x4 & 4x4 & 4x4 & 4x4 & 4x4 & 4x4 & 4x4 & 4x4 & 4x4 \\
       \hline
        Dilation & 1 & 1  & 2 & 3 & 5  &  & & & & &  \\ 
        \hline
        Mamba &  &   &  &  &   &  & & & & & Param$\#$ 16384  \\       
        \hline
        Transformer &  &   &  &  &   &  & & & & &  Param$\#$ 49152  \\       
        \hline
    \end{tabular}
    \vspace{-01.520em}
    \label{table:spectral enhancement network}
\end{table}

Mamba in the bottleneck layer captures the global correlation among consecutive phonemes from the spectrogram. Since Mamba is a specific sequence modeling architecture (originally for 1D sequences), we need to adapt it to work in the 2D bottleneck context. This can be done by flattening the spatial dimensions (H, W) into a sequence, applying Mamba, and then reshaping back. The Mamba is chosen over Transformers \cite{ashish2017attention} because Mamba requires one-third less parameters than Transformers (i.e., 16384 vs 49152) for the same embedding dimension (64) and sequence length (16 x 16) in our design (see Table \ref{table:spectral enhancement network}). 

\vspace{-0.5em}

\vspace{-01.72em}
\subsection{Upsampling network}
\label{subsec:Upsampling network}

The upsampling network is inspired by version 2 of HiFi-GAN \cite{kong2020hifi}, which takes the enriched spectrogram representation from the spectrum enhancement network as input and gives a raw waveform at its output. The input spectrogram for the upsampling network is a $\sim$1 second audio clip. Let's say, if the target sampling frequency is 16 kHz, the 1s clip will have 16000 samples. The hop size is 256 for the spectrogram. This means that there are 16000/256 $\approx$ 62 frames in the spectrograms. Therefore, the upsampling network needs to generate 16000 samples from 62 frames, which means the upsampling network needs $\sim$256x upsampling.

\begin{figure}[h!]
\vspace{-0.95em}
    \centering
    \includegraphics[width=0.495\textwidth,height=0.1\textheight]{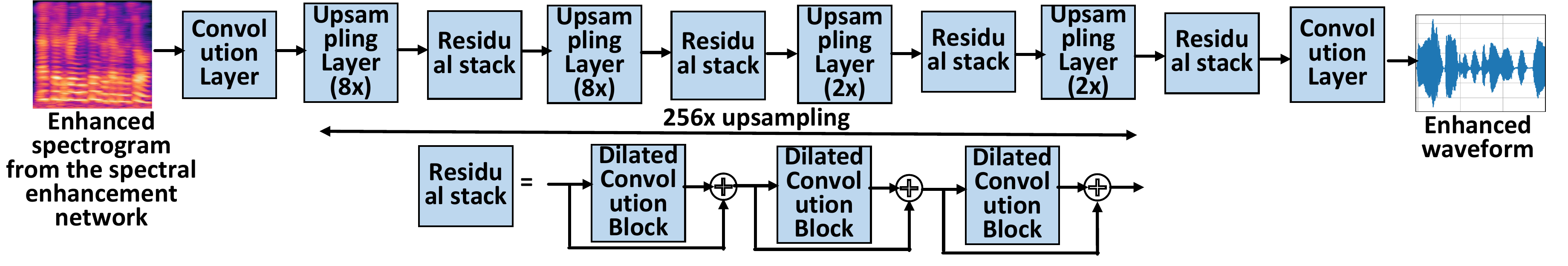}
    \vspace{-02.75em}
    \caption{The upsampling network has a 256 upsampling ratio, which is done in 4 stages: 8x, 8x, 2x, and 2x upsampling.}
    \label{fig:upsampling network}
    \vspace{-01.85em}
\end{figure}

Fig. \ref{fig:upsampling network} shows the upsampling network in detail. Each upsampling layer is a transposed convolution, where the kernel size is twice the stride. A stack of transposed convolutional layers is used to upsample the input sequence to 256x. The 256x upsampling is done in 4 stages: 8x, 8x, 2x, and 2x upsampling. Each transposed convolutional layer is followed by a residual layer. Each of the residual layers has three dilated convolutions with dilation 1, 3, and 9, with kernel size 3, having a total receptive field of 27 timesteps.  
The receptive field of a stack of dilated convolution layers increases exponentially with the number of layers. This effectively implies a larger overlap in the induced receptive field of far-apart time-steps, leading to better long-range correlation. Leaky ReLU activation is used in each residual layer after each dilated convolution. 

\vspace{-0.5em}
\subsection{Time enhancement network}
\label{subsec:time enhancement network}

\begin{figure}[h!]
\vspace{-0.975em}
    \centering
    \includegraphics[width=0.49\textwidth,height=0.15\textheight]{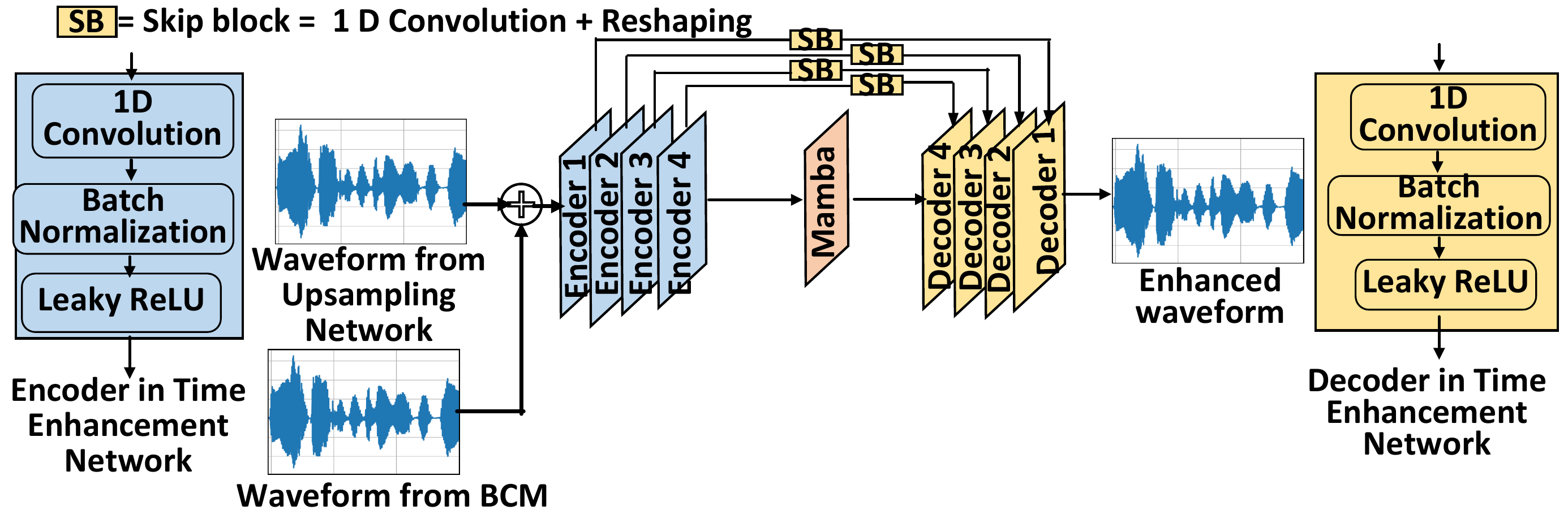}
    \vspace{-01.975em}
    \caption{The time enhancement network is designed to fuse features from both the acoustic (i.e., ACM) and vibration (i.e., BCM) modalities.} 
    \label{fig:time enhancement network}
    \vspace{-0.75em}
\end{figure}

The time enhancement network (see Fig. \ref{fig:time enhancement network}) is designed to fuse features from both the acoustic (i.e., ACM) and vibration (i.e., BCM) modalities. As signals from BCMs are less noisy compared to the signals from ACMs, the enhanced signals of ACMs from the spectrum enhancement network are further improved using the less noisy signal of BCMs by the time enhancement network. 

The time enhancement network is inspired by the well-known Wave-U-Net architecture \cite{stoller2018wave}, which is a fully convolutional 1D-UNet-like neural network. This network consists of a 1D convolution framework in U-Net, featuring 4 full 1D convolution layers as encoders, 4 full 1D convolution layers as decoders, and a Mamba \cite{gu2023mamba} block as a bottleneck layer (see Table \ref{table:time enhancement network} for details). Each 1D convolution layer incorporates batch normalization followed by leaky ReLU activations. 

\begin{table}[ht!]
\vspace{-0.951800em}
\scriptsize
\setlength{\tabcolsep}{2.1pt}
    \centering
    \centering
    \caption{The details of the convolution layers in the time enhancement network. Here, Enc = encoder and Dec = decoder.}
    \vspace{-0.651800em}
    \begin{tabular}{l | l|l|l|l | l|l|l|l | l}
    \hline
        \cellcolor [gray]{0.85}\textbf{Layer} & \cellcolor [gray]{0.85}\textbf{Enc1} & \cellcolor [gray]{0.85}\textbf{Enc2} & \cellcolor [gray]{0.85}\textbf{Enc3} & \cellcolor [gray]{0.85}\textbf{Enc4} &  \cellcolor [gray]{0.85}\textbf{Dec1} & \cellcolor [gray]{0.85}\textbf{Dec2} & \cellcolor [gray]{0.85}\textbf{Dec3} & \cellcolor [gray]{0.85}\textbf{Dec4} &  \cellcolor [gray]{0.85}\textbf{Bottleneck}\\ 
        \hline
        \hline
        Kernels & 10 &  20 & 40 & 80 & 80 & 40 & 20 & 10 &  \\ 
        \hline
       Kernel size & 4x4 & 4x4 & 4x4 & 4x4 & 4x4 & 4x4 & 4x4 & 4x4 & \\
       \hline
        Dilation & 1  & 2 & 3 & 5  &  & &  & &  \\ 
        \hline
        Mamba &  &   &  &  &   &  & & & Param$\#$ 25864  \\       
        \hline
        Transformer &  &   &  &  &   &  & & &  Param$\#$ 76895  \\       
        \hline
    \end{tabular}
    \vspace{-01.520em}
    \label{table:time enhancement network}
\end{table}

Since Mamba is a specific sequence modeling architecture originally for 1D sequences and the encoders/decoders process 1D sequences, we don't need to reshape Mamba. We compare Mamba with Transformers for the same embedding dimension = 80 and sequence length = 16 x 16 in our time enhancement network (see Table \ref{table:time enhancement network}). It shows that a roughly one-third reduction in parameter count happens for Mamba over Transformers (i.e., 25864 vs 76895). 


\vspace{-0.9520em}
\subsection{Amplitude-phase enhancement network}
\label{subsec:Spectral-phase enhancement network}


The purpose of this module is to remove artifacts and noise in both the \textit{amplitude and phase domains} from the output waveform in a learnable way. This module helps to generate a clean phase from the noisy phase for the successful reconstruction of audio signals in extremely noisy conditions (i.e., when both the ACMs and BCMs are noisy).

\begin{figure}[h!]
\vspace{-0.98em}
    \centering
    \includegraphics[width=0.49\textwidth,height=0.16\textheight]{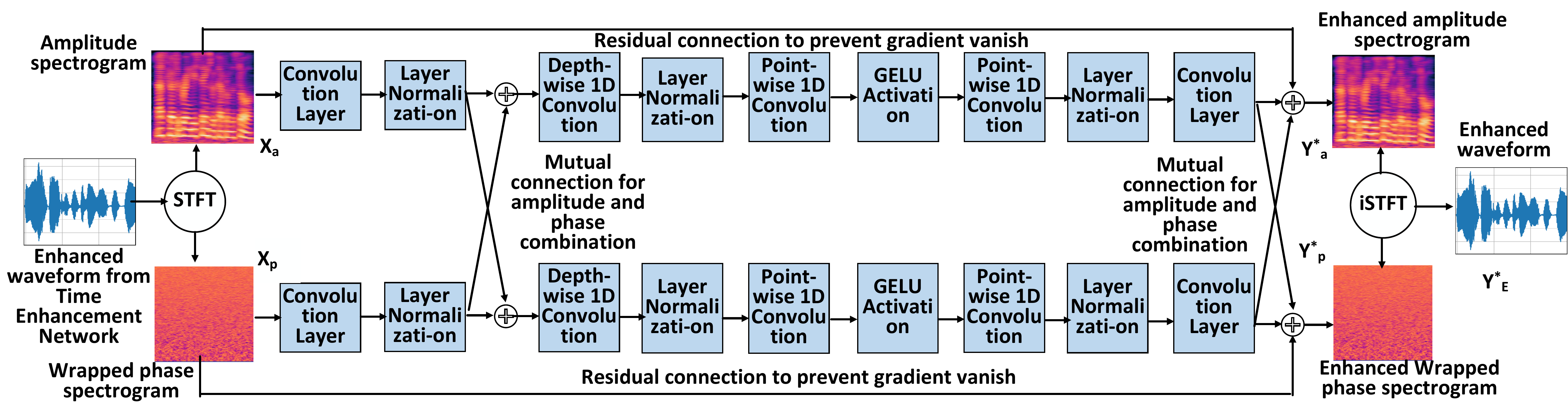}
    \vspace{-02.20em}
    \caption{The amplitude-phase enhancement network helps to generate a clean phase from the noisy phase for successful reconstruction of audio signals in noisy conditions (i.e., when both the ACMs and BCMs are noisy).}
    \label{fig:amplitude-phase enhancement network}
    \vspace{-0.8em}
\end{figure}

The network is shown in Fig. \ref{fig:amplitude-phase enhancement network}. The waveform from the time enhancement network is first subjected to a short-time Fourier transformation (STFT) that gives an amplitude spectrum $X_a \in \mathbb{R}^{T\times F}$ and a wrapped phase spectrum $X_p \in \mathbb{R}^{T\times F}$, where T = 125 and F = 513 denote the number of temporal frames and frequency bins, respectively. 
The amplitude $X_a$ and phase $X_p$ streams utilize an identical network of a series of 1D convolution operations with mutual coupling between the two streams. The 1D convolution operations contain a cascade of a large-kernel-sized depth-wise convolutional layer and a pair of point-wise convolutional layers that respectively expand and restore feature dimensions.  The depth-wise (DW) convolution is implemented using the Conv1D operation, and the point-wise convolution is implemented using linear layers (see Table \ref{table:amplitude-phase enhancement network} for details). Layer normalization \cite{ba2016layer} and Gaussian error linear unit (GELU) activation \cite{hendrycks2016gaussian} are interleaved between the layers. Finally, the residual connection is added before the output to prevent the gradient from vanishing.

The enhanced amplitude spectrum $Y^{*}_a$ and wrapped phase spectrum $Y^{*}_p$ are used together at the end of the amplitude-phase enhancement network to generate the enhanced waveform $Y^{*}_E$ using inverse STFT (iSTFT) as follows,  $Y^{*}_E = iSTFT(Y^{*}_a \cdot {\rm e}^{jY^{*}_p} )$.

\vspace{-0.951800em}
\begin{table}[ht!]
\footnotesize
    \centering
    \caption{The details of the amplitude-phase enhancement network.}
    \vspace{-0.51800em}
    \begin{tabular}{l | l|l}
    \hline
        \cellcolor [gray]{0.85}\textbf{Layer} & \cellcolor [gray]{0.85}\textbf{Conv1D} & \cellcolor [gray]{0.85}\textbf{DWConv1D}\\ 
        \hline
        \hline
        Kernels & 512 &  512   \\ 
        \hline
       Kernel size & 7x1 & 7x1  \\
       \hline
        Dilation & 1  & 1  \\ 
        \hline
        Padding & 1  & 3   \\ 
        \hline
    \end{tabular}
    \vspace{-0.20em}
    \label{table:amplitude-phase enhancement network}
\end{table}
\vspace{-0.600em}

\subsection{Loss Functions}
\label{subsec:Combined loss in the frequency, time, phase domains}


\begin{figure}[h!]
\vspace{-0.65em}
    \centering
    \includegraphics[width=0.49\textwidth,height=0.15\textheight]{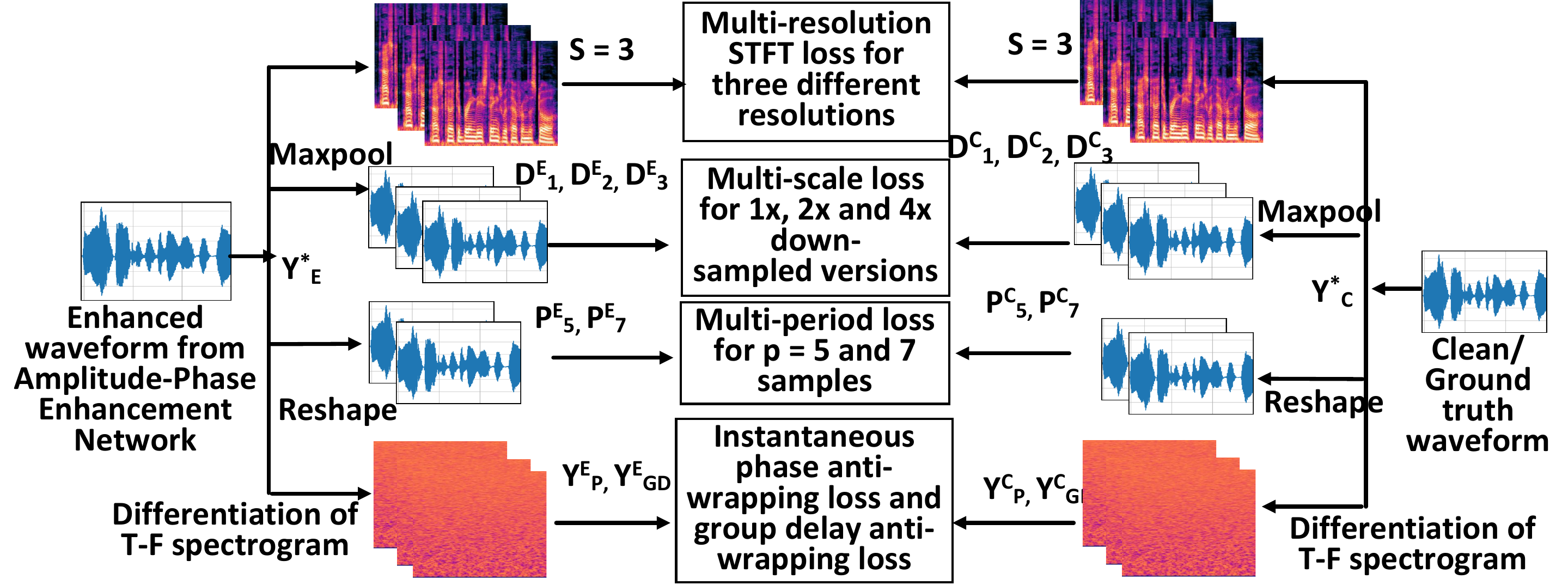}
    \vspace{-02.0em}
    \caption{The multi-scale and multi-period loss functions in SUBARU.}
    \label{fig:multiscalelossperiosloss}
    \vspace{-0.85em}
\end{figure}

\textbf{Multi-scale loss function:} 
Inspired by MelGAN \cite{kumar2019melgan}, SUBARU uses a multi-scale loss function shown in Fig. \ref{fig:multiscalelossperiosloss}. For that,  SUBARU generates three down-sampled audio $D^E_1, D^E_2,$ and $D^E_3$ using three max-pooling layers from the enhanced waveform $Y^{*}_E$ with 1x, 2x, and 4x downsampling ratio. Parallelly, SUBARU generates three down-sampled audio $D^C_1, D^C_2,$ and $D^C_3$ using three similar max-pooling layers from the clean/ground-truth waveform $Y^{*}_C$. Next, SUBARU calculates the mean absolute error (MAE) expressed by Eqn. \ref{eqn:multi-scale loss}.

\vspace{-1.0em}
{\scriptsize
\begin{equation}
\begin{aligned}
\text{Multi-scale loss} = \frac{1}{3\times N} \sum_{}^{N} |D^C_1 - D^E_1| +  |D^C_2 - D^E_2| + |D^C_3 - D^E_3|\\
\end{aligned}
\label{eqn:multi-scale loss}
\vspace{-0.6em}
\end{equation}
}


\textbf{Multi-period loss function:} 
Initially, SUBARU reshapes the enhanced waveform $Y^{*}_E$ (the output of the amplitude-phase enhancement network) from a 1D waveform into a 2D representation by segmenting it based on a specific period (p) of 5 and 7 samples. After reshaping, the dimension of the 2D tensor is (1, T / p, p), where T = number of audio samples and p = periods. 
Let us denote these two 2D tensors generated for p = 5 and 7 samples from the enhanced waveform $Y^{*}_E$ by $P^E_{5}$ and $P^E_{7}$, respectively. Similarly, SUBARU generates two 2D tensors for p = 5 and 7 from the clean/ground-truth waveform $Y^{*}_C$ and let us denote them by $P^C_{5}$ and $P^C_{7}$, respectively. SUBARU calculates the MAE energy loss between clean and enhanced audio at each period for 2D tensors (see 
 Eqn. \ref{eqn:multi-period loss}).


\vspace{-0.0em}
{\scriptsize
\begin{equation}
\begin{aligned}
\text{Multi-period loss} =  |\sum_{x,y}^{} P^C_5 - \sum_{x,y}^{} P^E_5| + |\sum_{x,y}^{} P^C_7 - \sum_{x,y}^{} P^E_7|\\
\label{eqn:multi-period loss}
\end{aligned}
\vspace{-0.5em}
\end{equation}
}
\vspace{-0.85em}



 SUBARU uses only 2 periods (i.e., p = 5 and 7) compared to 5 periods and 6 convolution layers in \cite{kong2020hifi} to keep the network size small without sacrificing audio quality.  

\textbf{Phase-spectrum loss function:} 
To reconstruct the clean phase when the BCMs are noisy, SUBARU includes a phase-spectrum loss function to enhance audio quality. Inspired by \cite{ai2023neural} and considering the phase wrapping issue, SUBARU proposes to use two anti-wrapping losses: one for the instantaneous phase and one for group delay. These two anti-wrapping losses use MAE loss as shown in Eq. \ref{eqn:antiwrapping}.

\vspace{-0.85em}
\vspace{-0.0em}
{\scriptsize
\begin{equation}
\begin{aligned}
\text{Instantaneous phase anti-wrapping loss} = \frac{1}{TF} \sum_{}^{TF} |f_{AW}(Y^C_P- Y^E_P)|\\
\text{Group-delay anti-wrapping loss} = \frac{1}{TF} \sum_{}^{TF} |f_{AW}(Y^C_{GD}- Y^E_{GD})|\\
\label{eqn:antiwrapping}
\end{aligned}
\vspace{-0.5em}
\end{equation}
}
\vspace{-01.95em}


Where $Y^C_P$ and $Y^C_{GD}$ are the instantaneous phase and group delay for the clean/ground-truth wave $Y^{*}_C$, respectively. The $Y^E_P$ and $Y^E_{GD}$ are the instantaneous phase and group delay for enhanced wave $Y^{*}_E$, respectively. The group delay $Y^C_{GD}$ and $Y^E_{GD}$ are calculated by taking differentiation along the frequency axis of the T-F spectrogram. The $f_{AW}(x)$ denotes the anti-wrapping function, which is defined as: $f_{AW}(x) = |x - 2\pi\cdot \text{round}(\frac{x}{2\pi})|, x \in \mathbb{R}$. 

\textbf{Multi-resolution STFT loss:} To improve the frequency content of the enhanced wave $Y^{*}_E$, SUBARU calculates the multi-resolution STFT loss \cite{tian2020tfgan} at S = 3 resolutions, such as \{frequency bins, hop sizes, window lengths\} = \{(256, 128, 256), (512, 256, 512), (1024, 512, 1024)\}, following Eq. \ref{eqn:mrstft}.

\vspace{-0.990em}
{\scriptsize
\begin{align}
\vspace{-0.5em}
  \text{Multi-resolution STFT loss} = \frac{1}{S} \sum_{s=1}^{S} \Big( L_{\mathrm{SC}}^{} + L_{\mathrm{mag}}^{} \Big)
  \label{eqn:mrstft}
\end{align}
}
\vspace{-0.95em}

Where, spectral convergence loss $L_{SC}$ \cite{tian2020tfgan} and log STFT magnitude loss $L_{mag}$ \cite{tian2020tfgan} are calculated for S= 3 resolutions.

\textbf{Total loss:} The total loss is the summation of multi-scale, multi-period, phase-spectrum, and multi-resolution STFT losses. The multi-scale and multi-period losses are time-domain loss functions. Therefore, SUBARU employs training in all three domains: time, phase, and frequency.

\vspace{-0.95em}
\subsection{Breakdown by size, parameters, and inference time}
\label{subsec:Breakdown of SUBARU size and parameters}




To facilitate the understanding of SUBARU for real-time streaming operations, we provide a breakdown of SUBARU by size, parameters, inference time, and floating-point operations (FLOPs) in Table \ref{table:breakdownmodel}. Inference time is measured on an NVIDIA 4090 GPU, an Intel(R) Xeon(R) Silver 4310 CPU (2.10 GHz), and Google Pixel7 for an audio frame of 1s.

\begin{table}[ht!]
\scriptsize
\setlength{\tabcolsep}{0.1pt}
\vspace{-0.91800em}
    \centering
    \caption{Breakdown by size, parameters, and inference time.}
    \vspace{-0.51800em}
    \begin{tabular}{l | l|l|l|l}
    \hline
        \cellcolor [gray]{0.85}\textbf{Network breakdown} & \cellcolor [gray]{0.85}\textbf{Parameter} & \cellcolor [gray]{0.85}\textbf{Size} & \cellcolor [gray]{0.85}\textbf{Inference(GPU/CPU/Pixel7)}  & \cellcolor [gray]{0.85}\textbf{FLOPs}\\ 
        \hline
        \hline
        Spectrum enhancement & 900.2 k & 3.42 MB & 0.41 ms / 5.6 ms / 16.5 ms & 3.15 G  \\ 
        \hline
       Upsampling & 1130.3 k & 4.30 MB & 0.68 ms / 10.12 ms / 27.5 ms & 3.95 G \\
       \hline
        Time enhancement & 760.5 k & 2.89 MB & 0.38 ms / 5.13 ms / 15.3 ms & 2.66 G \\ 
        \hline
        Amplitude-phase & 828.6 k  &  3.15 MB & 0.27 ms / 3.65 ms / 10.92  ms & 2.90 G\\ 
        \hline
        \textbf{Total} & \textbf{3.61 M} & \textbf{13.77 MB} & \textbf{1.74 ms / 24.54 ms / 70.41 ms } & \textbf{12.67 G}\\
         \hline
    \end{tabular}
    \vspace{-01.820em}
    \label{table:breakdownmodel}
\end{table}

\subsection{Streaming operation of SUBARU}
\label{subsec:Streaming operation of SUBARU}


 The low-resolution audio sampled at the sub-Nyquist frequency in hearables is transmitted over Bluetooth to the mobile platform for BWE and multimodal SE. The low-resolution audio transmitted from the hearables are received by the mobile platforms frame by frame. For real-time streaming applications, the inference time of each frame must be less than the frame duration. From Table \ref{table:breakdownmodel}, it is clear that SUBARU's inference time for both GPUs, CPUs, and Pixel7 is much shorter (i.e., 1.74 ms/24.54 ms/70.41 ms) than the frame duration of 1s. Therefore, SUBARU is suitable for streaming audio from hearables to mobile platforms. 

\begin{figure}[h!]
\vspace{-0.85em}
    \centering
    \includegraphics[width=0.49\textwidth,height=0.15\textheight]{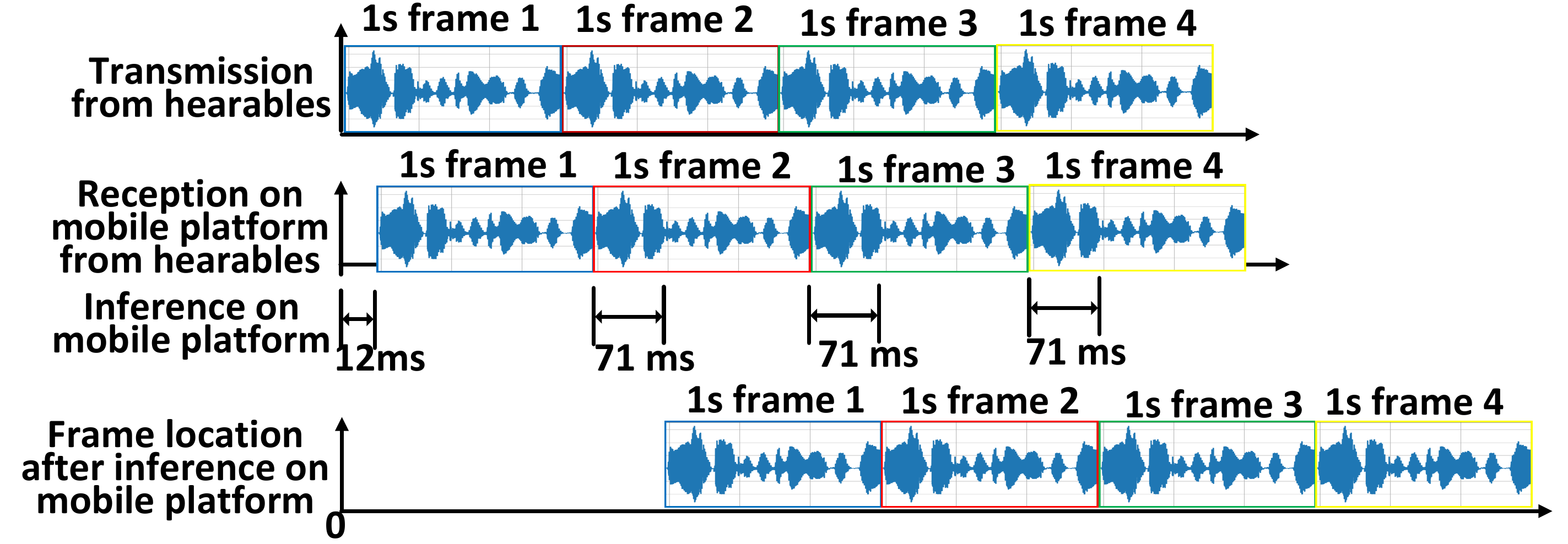}
    \vspace{-02.1em}
    \caption{The frame-by-frame real-time streaming operation of SUBARU.}
    \label{fig:framestream}
    \vspace{-0.65em}
\end{figure}

In Fig. \ref{fig:framestream}, at t = 0, the clock starts and the low-resolution audio from hearables starts to be transmitted to the mobile platform (i.e., Google Pixel7). We use the NRF52840 chip with the Bluetooth sniffer to emulate the streaming operation of SUBARU. At t = 12 ms, low-resolution audio packets are received and unpacked on the mobile platform. Next, SUBARU starts audio reconstruction and takes around 71 ms to finish the reconstruction. Therefore, there is a 12 + 71 = 83 ms latency for each 1s audio frame for the end-to-end BWE +  SE operation from hearables to mobile platforms. Please note that the 71 ms delay is for the actual model inference, and the 12 ms delay within the 83 ms total latency is not related to SUBARU, but rather related to the time required for audio frame packing, transmission, reception, and unpacking. 
Moreover, similar latency exists in all the commercial hearables, such as Apple AirPods Pro (1st \& 2nd Gen) has a latency of  $\sim$120 ms \cite{coyle_airpods_latency_2022}, and Samsung Galaxy Buds Pro has a latency of  $\sim$60 ms \cite{rtings_galaxy_buds_2021}.  For real-world audio communication, one-way delays up to 150 ms are considered acceptable for most user applications, including voice calls, according to the recommendation of the International Telecommunication Union (ITU) G.114 \cite{ITU-G114}. 
Please also note that, as the inference time is shorter than 1s frame duration, the frames are always streaming in real-time. 

\vspace{-0.65em}
\section{IMPLEMENTATION}
\label{sec:implementation}


\begin{figure}[h!]
\vspace{-0.65em}
    \centering
    \includegraphics[width=0.45\textwidth,height=0.14\textheight]{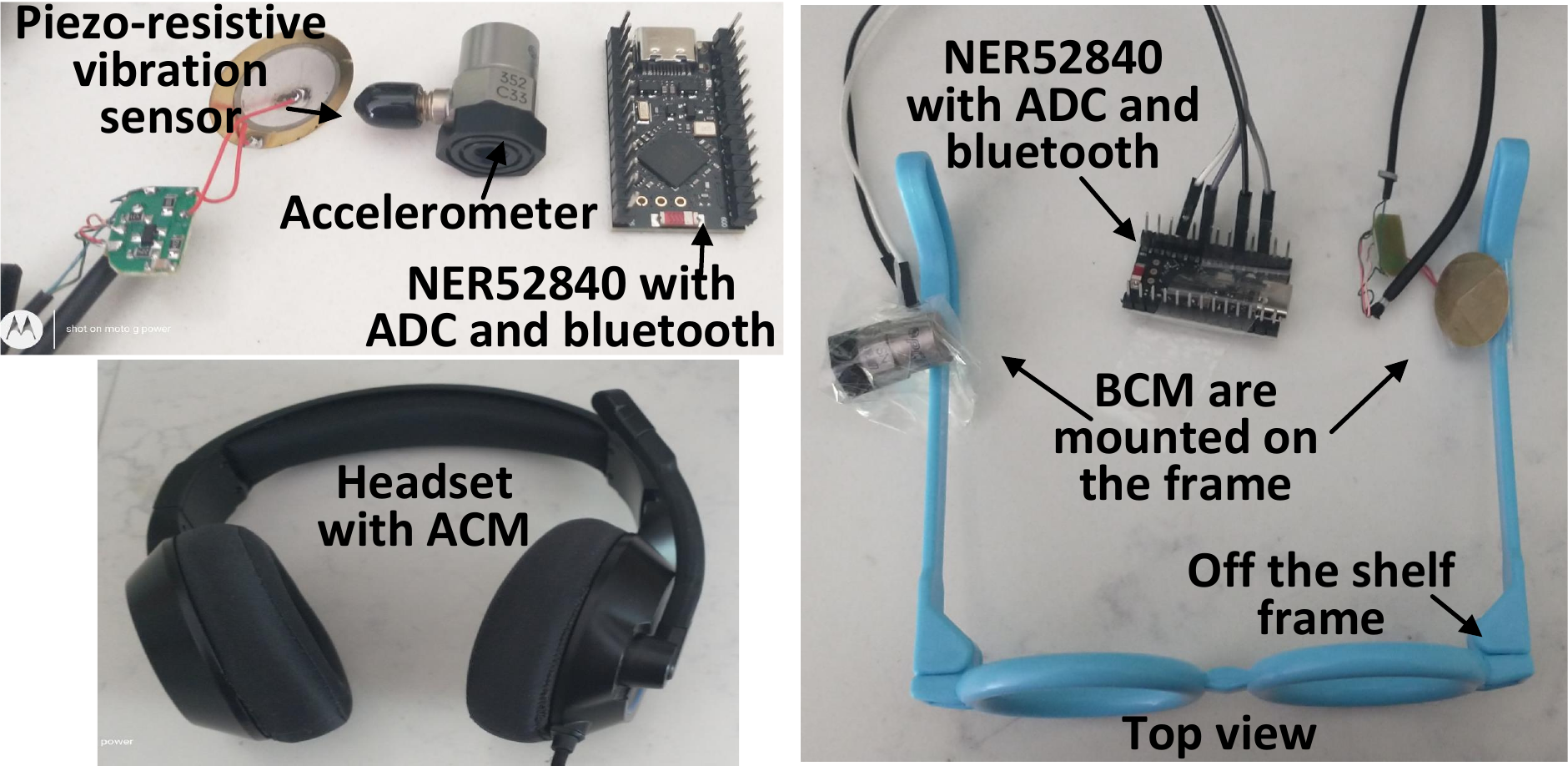}
    \vspace{-01.0em}
    \caption{A prototype of hearable. An off-the-shelf vibration sensor and an accelerometer are used as BCMs. Built-in ADC from NRF52840 samples signals from ACMs and BCMs at different Nyquist sampling frequencies and low-bit resolutions.}
    \label{fig:hardwaresetup}
    \vspace{-1em}
\end{figure}

\vspace{-0.85em}
\subsection{Wearable platform and mobile platform design}
\label{subsec:Hardware design}

We use two analog sensors for BCMs to prove and verify the strength of SUBARU: one is a piezo-resistive vibration sensor (part\# CEB-27032-L100) \cite{ceb27032l100} and another one is an accelerometer (part\# 352C33) \cite{pcb352c33}. 
The vibration sensor and the accelerometer are attached with an off-the-shelf plastic frame for collecting vibration near the earbone (see Fig. \ref{fig:hardwaresetup}).  As both are analog sensors, the analog output from these two sensors is sampled and transmitted by the NRF52840 chip over Bluetooth to a mobile platform. We use the built-in ADC from the NRF52840 chip to sample the analog signals from the BCMs and transmit over Bluetooth to a mobile platform. This helps us to control the sampling frequencies from the sub-Nyquist to Nyquist range and ADC bit resolution (8 to 12 bits) while sampling from BCMs.  We vary the sampling frequency from 4 kHz to 22 kHz for BCMs and also for ACMs. We use an off-the-shelf microphone from NUbwo as an ACM.

As our low-resolution audio sampled from BCMs will go through joint BWE and multimodal SE in a mobile platform, we use two devices as mobile platforms to compare performance: one is a computer with a 4090 GPU with Intel(R) Xeon(R) Silver 4310 CPU, and another one is a Google Pixel7 having Google Tensor G2 chipset and Mali-G710 MP7 GPUs. 


\vspace{-1.5em}
\subsection{Dataset collection}
\label{subsec:Dataset collection}


We collect an in-house dataset with our ACMs and BCMs mounted near the ear bone. Although there are a few datasets \cite{hauret2024vibravox, wang2022end} available for multimodal SE, those datasets use only one BCM. Therefore, they don't have simultaneous data from both vibration and accelerometer sensors with microphones at different ADC bit resolutions and at different noisy conditions. Therefore, we ask 20 persons (11 males and 9 females) with consent for data collection from the ACM and the two BCMs simultaneously at a 22 kHz sampling frequency. We select text from the VCTK dataset \cite{yamagishi2019cstr}, which contains a variety of dialogues. For this purpose, we convert the VCTK audio to text using Whisper \cite{whisper2022}.  We apply a high-pass filter with a cut-off frequency of 15 Hz to remove any body movement from the collected data. The data is then normalized and clipped within -1 to +1. We collect data from each person for 10 minutes in a quiet room (no noise)  for 3 different bit resolutions - 12, 10, and 8 bits - for the two different BCMs.

As the collected data amount for \{ACM, BCM\} pair is not sufficient, we create a synthetic dataset for better model training. To do this, we first need to model the two BCMs (vibration sensor and accelerometers) for three different bit resolutions of 12, 10, and 8 bits. We choose SEANet \cite{tagliasacchi2020seanet} for this purpose. We train six different SEANet models for six different input-target pairs using our in-house data, shown in Table \ref{table:syntheticdata}. Then, we use the six trained SEANets to generate synthetic data for six \{ACM, BCM\} pairs by feeding samples from the VCTK dataset.

\begin{table}[ht!]
\vspace{-0.91800em}
\scriptsize
\setlength{\tabcolsep}{3.5pt}
    \centering
    \caption{Modeling both BCMs - vibration and accelerometer for generating synthetic data.}
    \vspace{-0.51800em}
    \begin{tabular}{l | l|l}
    \hline
        \cellcolor [gray]{0.85}\textbf{Name} & \cellcolor [gray]{0.85}\textbf{Input in-house data} & \cellcolor [gray]{0.85}\textbf{Target in-house data}\\ 
        \hline
        \hline
        $SEANet_{12V}$ & 12-bit audio from ACM &  12-bit audio from vibration sensor   \\ 
        \hline
       $SEANet_{8V}$ & 8 bit audio from ACM &  8 bit audio from vibration sensor    \\
        \hline
       $SEANet_{10V}$ & 10 bit audio from ACM &  10 bit audio from vibration sensor    \\
       \hline
        $SEANet_{12A}$ & 12-bit audio from ACM &  12-bit audio from accelerometer    \\ 
        \hline
        $SEANet_{8A}$ & 8 bit audio from ACM &  8 bit audio from accelerometer    \\ 
        \hline
        $SEANet_{10A}$ & 10 bit audio from ACM &  10 bit audio from accelerometer    \\ 
        \hline
    \end{tabular}
    \vspace{-01.20em}
    \label{table:syntheticdata}
\end{table}

Two noise sources are used: non-speech noise and speech noise. For non-speech noises, we choose from a diverse set of noise types from \cite{font2013freesound} and randomly mix with the clean audio from the ACM. For speech noises,  we employ speech samples from Librispeech from different speakers.




\vspace{-0.90em}
\subsection{Model training}
\label{subsec:Model training}


The training is carried out in an end-to-end fashion. The noisy audio from the ACM for 12, 10 and 8-bit resolutions is given together as input to the model in the time domain. The signals from either of the BCMs are also given as an input in the time domain. For training our proposed SUBARU, all the audio clips underwent silence trimming and were sliced into $\sim1s$ clips. 
Since the synthetic data inevitably slightly differs from the real data, incorporating all the synthetic data together into a batch may degrade the model performance. To prevent the model from degrading during the training phase, we use an equal proportion of real and synthetic data in the initial training epochs. As training advances, the ratio of synthetic data is progressively reduced, eventually transitioning to batches comprised solely of real data.

The key training parameters include a batch size of 8 with $\sim$50 epochs, and the Adam optimizer with a learning rate of \(1\times10^{-4}\), weight decay of \(1\times10^{-5}\), and momentum parameters \(\beta_1=0.5\) and \(\beta_2=0.999\). The learning rate is scheduled using cosine annealing warm restarts (with \(T_0=10\) and \(T_{\text{mult}}=1\)), gradient clipping (max norm of 10), and gradient accumulation (over 2 batches) to ensure stability. Training is first executed on a computer with a 4090 GPU with Intel(R) Xeon(R) Silver 4310 CPU. The trained model is then customized to run on a Google Pixel7 smartphone for inference (see Section \ref{sec:performance evaluation}).

\vspace{-0.90em}
\subsection{Evaluation metrics}
\label{subsec:Evaluation Metrics}

To comprehensively evaluate in terms of intelligibility, fidelity, and perceived quality, we use Log-Spectral Distance (LSD), Short-Time Objective Intelligibility (STOI), Perceptual Evaluation of Speech Quality (PESQ), Scale-Invariant Signal-to-Distortion Ratio (SI-SDR), Non-Intrusive Speech Quality Assessment - Mean Opinion Score (NISQA-MOS), and Virtual Speech Quality Objective Listener (VISQOL).

\vspace{-0.90em}
\subsection{Base models}
\label{subsec:Base models}

We compare ten base models from Table \ref{table:summaryjointBWE} with our proposed SUBARU. Four (ATS-UNet, TFiLM, AFiLM, VibVoice) of the models are pure U-Net based, four (SEANet, AERO, EBEN, HiFi++) are pure GAN based, one (NU-Wave) is a diffusion probabilistic model, and one (NVSR) has U-Net with GAN vocoder. We cannot compare with TRAMBA as it does not have open-source code available.

\vspace{-0.70em}
\section{Performance evaluation}
\label{sec:performance evaluation}

This section evaluates the performance of SUBARU on two different platforms: desktop and Google Pixel7 smartphone. The details of both platforms are  explained in Section \ref{subsec:Hardware design}.

\textbf{Preparing to evaluate on Google Pixel7:} We can evaluate the model on the desktop platform in a straightforward manner. However, to evaluate the model on the Google Pixel7, we need to do some extra steps. The training is first done in PyTorch on the desktop following Sections \ref{subsec:Dataset collection}, and \ref{subsec:Model training}. The trained model in PyTorch is then exported to the ONNX model and then converted to TensorFlow Lite (TFLite) via ONNX-TensorFlow \cite{onnx_tf} on the desktop. Then the TFLite model is integrated into the TensorFlow Lite Android API. Then we run the TFLite model using TFLite GPU Delegate on the Google Pixel7 for inference. The TFLite GPU Delegate is used to utilize the Pixel's Mali GPU for faster inference.

\vspace{-0.90em}
\subsection{Evaluation for speech enhancement with inference time}
\label{subsec:Evaluation for joint BWE and multi-modal SE}

We select four SOTA GAN models (SEANet, AERO, HiFi++, and EBEN) and two U-Net models (VibVoice, TFiLM),  for speech enhancement evaluation. Please note that VibVoice and SEANet are by default multimodal SE models, whereas TFiLM, AERO, EBEN, and HiFi++ are single-modal speech enhancement models. To compare our model with the single-modal SE models, we convert our multimodal  SUBARU into a single-modal model by making it a single-input network (i.e., we don't feed the BCM signal to the time enhancement network). We name the single-modal version of SUBARU as SUBARU-single. We train all multimodal models using noisy audio from both ACMs and BCMs, and all single-modal models using noisy audio from the ACM only. 
To diversify the evaluation, we also introduce a music dataset MagnaTagATune \cite{law2009evaluation}. The results are shown in Table \ref{table:evaluationforSE} for 4-16 kHz BWE with noisy data having noise in between -7 to 5 dB with 12-bit ADC resolution for the vibration sensor on \textit{the desktop only.} 

Table \ref{table:evaluationforSE} indicates that  SUBARU is superior to the SOTA U-Net models for both the speech and music datasets. Moreover, SUBARU achieves better performance with the lowest inference time (i.e., 1.74 ms) compared to the high-resource GAN models in noisy conditions. Specifically, SUBARU needs 3.61x less inference time compared to HiFi++  and 20.68x less than AERO (i.e., the two best-performing models). This indicates that SUBARU  achieves better joint BWE and SE under noisy conditions without sacrificing the perceptual quality and intelligibility of the audio compared to the GAN counterpart.  

Please note that SUBARU is the smallest model with 20x fewer parameters than HiFi++ GAN and 10x fewer parameters than AERO. The parameter number of the GAN models includes both the generators and discriminators. 

\begin{table}[ht!]
\vspace{-01.51800em}
    \scriptsize
\setlength{\tabcolsep}{1.3pt}
    \centering
    \caption{Evaluating SE for 12-bit resolutions on the desktop for 4 - 16 kHz upsampling with noisy data for the vibration sensor. Here, L = LSD, V = VISQOL, N = NISQA-MOS, S = SI-SDR, P = PESQ, ST = STOI, Param = Parameter and Infer. = Inference.}
    \vspace{-0.51800em}
    \begin{tabular}
     {m{01.4cm}|m{0.7 cm}|m{0.7 cm}|m{0.6 cm}|c|m{0.5 cm}|c|c|c|c|c|c}
    \hline

         \cellcolor [gray]{0.85}\textbf{Model} & \cellcolor [gray]{0.85}\textbf{Type} & \cellcolor [gray]{0.85}\textbf{Param (M)} & \cellcolor [gray]{0.85}\textbf{Size (MB)} & \cellcolor [gray]{0.85}\textbf{Dataset}  & \cellcolor [gray]{0.85}\textbf{Infer. (ms)}  & \cellcolor [gray]{0.85}\textbf{L $\downarrow$} & \cellcolor [gray]{0.85}\textbf{V $\uparrow$} & \cellcolor [gray]{0.85}\textbf{N $\uparrow$} & \cellcolor [gray]{0.85}\textbf{S $\uparrow$} & \cellcolor [gray]{0.85}\textbf{P $\uparrow$} & \cellcolor [gray]{0.85}\textbf{ST $\uparrow$} \\ 
         \hline
         
        \hline
        \hline
         \multirow{2}{*} {Unprocessed}  &  & &  & VCTK &   &  2.78 & 1.84  & 1.27 & 8.54 & 1.11 & 0.79 \\
         &  & &  & Magna &   &  2.85 &  1.62  & 1.21 & 7.28 & 1.03 & 0.78  \\
        
        \Xhline{3\arrayrulewidth}

         \multirow{2}{*} {TFiLM \cite{birnbaum2019temporal}} & \multirow{2}{*} {U-Net} & \multirow{2}{*} {68.2} & \multirow{2}{*} {260.3} & VCTK & 4.85  & 1.68    &  3.73    &  3.53   &  10.28   &  2.03 &  0.81  \\ 
          & &  &  & Magna &  4.85 &  1.70   &  3.67    & 3.46    &  9.41    &  1.99 &  0.81   \\ 
\hline

\multirow{2}{*} {VibVoice \cite{he2023vibvoice}} & \multirow{2}{*} {U-Net} & \multirow{2}{*} {23.14} & \multirow{2}{*} {5.8 }& VCTK & 17.2  & 3.1  &  2.73    & 2.51    &  12.48    &  2.05 &  0.81  \\ 
         & &  &   & Magna & 17.2  & 3.28   &  2.66    & 2.43    &   11.21  &  2.01  &   0.80   \\ 

        \Xhline{3\arrayrulewidth}   
          
         \multirow{2}{*} {AERO \cite{mandel2023aero}} & \multirow{2}{*} {GAN} & \multirow{2}{*} {36.3} & \multirow{2}{*} {138.7} &  VCTK & 36 &  0.97  &  4.16   &  4.03    &  17.03 &  2.93   &   0.89  \\ 
         & &  &  & Magna &  36 & 0.99    &  4.10   &  3.98    &   16.29  &  2.89  &   0.88  \\ 
          
         \hline
        \multirow{2}{*} {EBEN \cite{hauret2023eben}} & \multirow{2}{*} {GAN} & \multirow{2}{*} {29.7} & \multirow{2}{*} {113.3} & VCTK  & 12.7 &  1.15 &   3.78   & 3.65    & 14.23  & 2.57  & 0.84  \\ 
         & & &  &  Magna  & 12.7 & 1.17    &   3.76    & 3.63    & 13.13  & 2.54 & 0.88   \\ 
           
         \hline
        \multirow{2}{*} {HiFi++ \cite{kim2023hifi++}} & \multirow{2}{*} {GAN} & \multirow{2}{*} {72.2} & \multirow{2}{*} {259.92} &  VCTK  &  6.3 & 0.89   &   4.18   &  4.11   & 17.48   &  2.85 &  0.90  \\ 
         & & & &  Magna &  6.3 & 0.91   &   4.13   &  4.07   & 16.65   & 2.83   &   0.89  \\ 
          
         \hline
         \multirow{2}{*} {SEANet \cite{tagliasacchi2020seanet}}   &  \multirow{2}{*} {GAN} & \multirow{2}{*} {64.9} & \multirow{2}{*} {240.13} & VCTK & 9.18 & 1.39  &   3.89  &   3.78   &  14.31  & 2.43 &   0.89  \\
         & & & &  Magna  & 9.18 & 1.44   &  3.79    &   3.67    & 13.74     & 2.40 &  0.87   \\
         
        \hline
        \hline
       \multirow{2}{*} {\textbf{SUBARU}}  &  \multirow{2}{*} {U-Net} & \multirow{2}{*} {3.61}  & \multirow{2}{*} {13.77} & VCTK & \textbf{1.74} & \textbf{0.84} &  \textbf{4.35}  &   \textbf{4.19} &   \textbf{17.94}    &  \textbf{3.00} & \textbf{0.90}  \\
       & & &  &  Magna &  \textbf{1.74}  & \textbf{0.86} &  \textbf{4.27}  &   \textbf{4.11} &   \textbf{16.71}  &  \textbf{2.95}  &  \textbf{0.90} \\

          \multirow{2}{*} {\vspace{3mm}\textbf{SUBARU-}} & \multirow{2}{*} {U-Net} & \multirow{2}{*} {3.61} & \multirow{2}{*} {13.77} &  VCTK & \textbf{1.74}  & \textbf{0.86} &  \textbf{4.19}  &   \textbf{4.17} &   \textbf{17.49}    &  \textbf{2.97} & \textbf{0.90}   \\
       \textbf{single} & &  & &  Magna &  \textbf{1.74}  & \textbf{0.87} &  \textbf{4.14}  &   \textbf{4.09} &   \textbf{16.71}  &  \textbf{2.95}  &  \textbf{0.90} \\
        \hline
    \end{tabular}
    \vspace{-0.20em}
    \label{table:evaluationforSE}
    \vspace{-0.5800em}
\end{table}

\begin{table}[ht!]
\vspace{-01.51800em}
    \scriptsize
\setlength{\tabcolsep}{0.41pt}
    \centering
    \caption{Evaluating SE for 12-bit resolutions on the desktop and Google Pixel7 platforms for 4 - 16 kHz upsampling with noisy data for both the vibration sensor and the accelerometer.}
    \vspace{-0.51800em}
    \begin{tabular}
     {l|l|l|l|l|l|l|l|l||l|l|l|l|l|l}
    \hline

         \cellcolor [gray]{0.85}\textbf{Model}  & \cellcolor [gray]{0.85}\textbf{Device} & \cellcolor [gray]{0.85}\textbf{Infer.} & \multicolumn{6}{c||} {\cellcolor [gray]{0.85}\textbf{Vibration sensor}} &  \multicolumn{6}{|c} {\cellcolor [gray]{0.85}\textbf{Accelerometer}} \\ 
         \hline
          &  & (ms)  &  \textbf{L$\downarrow$} & \textbf{V$\uparrow$}  & \textbf{N$\uparrow$} & \textbf{S$\uparrow$}  & \textbf{P$\uparrow$} & \textbf{ST$\uparrow$} & \textbf{L$\downarrow$} & \textbf{V$\uparrow$}  & \textbf{N$\uparrow$} & \textbf{S$\uparrow$}  & \textbf{P$\uparrow$} & \textbf{ST$\uparrow$}\\
         
        \hline
        \hline
         \multirow{2}{*} {Unprocessed}  &  Desktop &  & 2.78   & 1.84 & 1.27 & 8.54 & 1.11 & 0.71 & 2.95   & 1.57 & 1.01 & 7.68 & 1.03 & 0.69  \\
                               & Pixel7  &  &  2.79   & 1.83 & 1.26 & 8.52 & 1.10 & 0.70 &   2.96   & 1.56 & 1.00 & 7.65 & 1.02 & 0.69  \\
        
        \Xhline{3\arrayrulewidth}

         \multirow{2}{*} {\vspace{3mm}TFiLM\cite{birnbaum2019temporal}}   & Desktop & 4.85 & 1.68   &  3.73   &  3.53   &  10.28   &  2.03 &  0.81     & 1.82   &  3.47    & 3.36    &  9.58    &  1.91 &  0.79  \\ 
           (U-Net)  & Pixel7  & 198 & 1.69   &  3.70     & 3.50   &  10.19   &  2.00 &  0.81   &  1.83  &  3.45   & 3.34   &   9.52  &  1.88  &   0.79  \\ 
\hline

\multirow{2}{*} {\vspace{3mm}VibVoice\cite{he2023vibvoice}}  & Desktop & 17.2  & 3.1  &  2.73    & 2.51    &  12.48    &  2.05 &  0.81  &  3.5   &  2.54    & 2.41    &  11.43    &  1.93 &  0.80  \\ 
          (U-Net) &  Pixel7  & 707  & 3.11   &  2.70    & 2.47    &   12.45  &  2.01  &   0.80  &  3.48   &  2.51    & 2.38   &   11.39  &  1.90  &   0.79  \\ 

        \Xhline{3\arrayrulewidth}   
          
         \multirow{2}{*} {\vspace{3mm}AERO\cite{mandel2023aero}}  & Desktop & 36 & 0.97  &  4.16   &  4.03    &  17.03 &  2.93   &   0.89 &  1.03   &  4.04   &  3.91    &  16.65 &  3.86   &   0.90 \\ 
          (GAN)  & Pixel7  & 1441 &  0.98    &  4.14   &  4.01    &   16.97  &  2.89  &   0.88 &  1.05    &  4.01   &  3.89    &   16.57  &  3.84  &   0.90  \\ 
          
         \hline
        \multirow{2}{*} {\vspace{3mm}EBEN\cite{hauret2023eben}}  & Desktop & 12.7& 1.15 &   3.78   & 3.65    & 14.23  & 2.57  & 0.84  &  1.21     &   3.58  & 3.37    & 13.27  & 2.39 & 0.83  \\ 
         (GAN)   & Pixel7  & 520 & 1.17    &   3.76    & 3.63    & 14.19  & 2.56 & 0.84 &  1.22   &   3.54    &  3.33   & 13.21  & 2.35 & 0.83  \\ 
           
         \hline
        \multirow{2}{*} {\vspace{3mm}HiFi++\cite{kim2023hifi++}}  & Desktop & 6.3 & 0.89   &   4.18  & 4.11   & 17.48   &  2.85 &  0.90 &   0.92   &   4.10   &  4.03   & 16.87   & 2.81 &  0.90 \\ 
         (GAN) & Pixel7  & 258 & 0.91   &   4.15   &  4.09   & 17.43   & 2.83   &   0.90 &  0.94   &   4.07   &  3.98   & 16.71   & 2.78   &   0.90   \\ 
          
         \hline
         \multirow{2}{*} {\vspace{3mm}SEANet\cite{tagliasacchi2020seanet}}    & Desktop & 9.18 & 1.39  &   3.89  &   3.78   &  14.31  & 2.43 &   0.89  & 1.51   &  3.67    &   3.42   & 13.24    & 2.31 &   0.87  \\
         (GAN)  & Pixel7  & 380  & 1.40   &  3.85    &   3.75    & 14.24     & 2.40 &  0.88 & 1.52   &  3.62    &   3.40    &  13.18     & 2.28 &  0.87   \\
         
        \hline
        \hline
       \multirow{2}{*} {\textbf{SUBARU}}   & Desktop & \textbf{1.74} & \textbf{0.84} &  \textbf{4.35}  &   \textbf{4.19} &   \textbf{17.94}    &  \textbf{3.00} & \textbf{0.90}  &  \textbf{0.87} &  \textbf{4.28}  &   \textbf{4.14} &   \textbf{17.05}    &  \textbf 2.95 & \textbf{0.89}  \\
           & Pixel7  & \textbf{70.41} & \textbf{0.85} &  \textbf{4.31}  &   \textbf{4.15} &   \textbf{17.71}  &  \textbf{2.97}  &  \textbf{0.90} &  \textbf{0.87} &  \textbf{4.25}  &   \textbf{4.11} &   \textbf{16.78}  &  \textbf{2.91}  &  \textbf{0.89} \\

          \multirow{2}{*} {\vspace{3mm}\textbf{SUBARU-}}   & Desktop &  \textbf{1.74} & \textbf{0.86} &  \textbf{4.19}  &   \textbf{4.17} &   \textbf{17.49}    &  \textbf{2.97} & \textbf{0.90}  &  \textbf{0.89} &  \textbf{4.15}  &   \textbf{4.10} &   \textbf{16.97}    &  \textbf{2.92} & \textbf{0.89}  \\
           \textbf{single} & Pixel7  & \textbf{70.41} & \textbf{0.86} &  \textbf{4.14}  &   \textbf{4.14} &   \textbf{17.31}  &  \textbf{2.95}  &  \textbf{0.90} &  \textbf{0.90} &  \textbf{4.11}  &   \textbf{4.05} &   \textbf{16.58}  &  \textbf{2.87}  &  \textbf{0.89} \\
        \hline
    \end{tabular}
    \vspace{-0.20em}
    \label{table:evaluationforSEforvibration and accel}
    \vspace{-0.5800em}
\end{table}

We further compare our SUBARU for accelerometers on Google Pixel7, and the results are shown in Table \ref{table:evaluationforSEforvibration and accel}. To maintain simplicity, we only consider the VCTK dataset. Table \ref{table:evaluationforSEforvibration and accel} indicates that the vibration sensor provides better results than the accelerometer at hand. The possible reason may be that the accelerometer is noisier than the vibration sensors.

\begin{figure}[h!]
\vspace{-0.0em}
    \centering
    \includegraphics[width=0.49\textwidth,height=0.18\textheight]{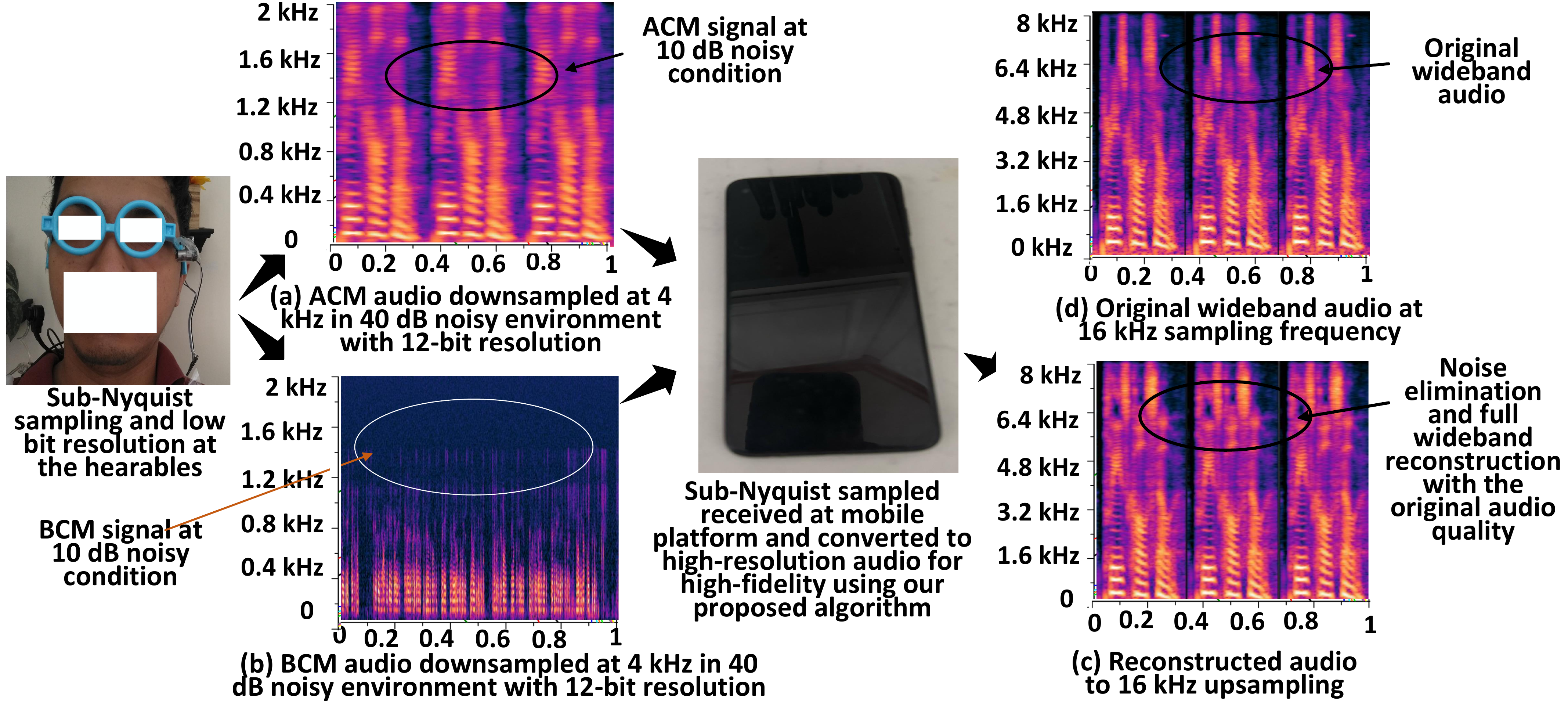}
    \vspace{-02.1em}
    \caption{Demonstration of full wideband reconstructed audio from the sub-Nyquist sampled and low-resolution audio at noisy conditions for 4-16 kHz. 
    The reconstructed audio is close to the original audio in terms of all metrics.}
    \label{fig:evaluation picture}
    \vspace{-01.5em}
\end{figure}

\vspace{-0.85em}
\subsection{Evaluation of real-time processing on desktop and Pixel7}
\label{subsec:inference speed at desktop and Google Pixel7}

We evaluate SUBARU on both desktop and Google Pixel7 platforms for a better understanding of the real-time behavior of SUBARU, which is critical for streaming operation of sound from the hearables. The results are summarized in Table \ref{table:evaluationforSEforvibration and accel}. 
We keep the model quantization the same for both desktop and Pixel7 while evaluating the inference time. All the models in Table \ref{table:evaluationforSEforvibration and accel} are suitable for real-time operation on the desktop platform, as the inference time is lower than the audio frame (1s). 
Moreover, for the streaming operation in audio communication, one-way delays up to 150 ms are considered acceptable for most user applications, including voice calls, according to the recommendation of the International Telecommunication Union (ITU) G.114 \cite{ITU-G114}. Therefore, all the models in Table \ref{table:evaluationforSEforvibration and accel} except SUBARU are not suitable for streaming operation on the Pixel7 platform as they have an inference time higher than 150 ms. \textit{In contrast, our proposed SUBARU has inference time smaller than 150 ms on both desktop and Pixel7, indicating its capability of streaming operation on both desktop and smartphone platforms.}

\begin{table}[h!]
\vspace{-01.400em}
\scriptsize
\setlength{\tabcolsep}{1.3pt}
\centering
\caption{NRF52840 ADC-only power consumption at different sampling rates and resolutions.}
\vspace{-0.51800em}
\begin{tabular}{c|c|c|c}
\hline
 \cellcolor [gray]{0.85}\textbf{Sampling Rate (Hz)} &  \cellcolor [gray]{0.85}\textbf{Resolution (bits)} &  \cellcolor [gray]{0.85}\textbf{Current draw (µA)} &  \cellcolor [gray]{0.85}\textbf{Power (mW) @ 3.0 V} \\
\hline
\multirow{3}{*}{4 kHz} & 8-bit  & 234  & 0.702  \\
                       & 10-bit & 248  & 0.744  \\
                       & 12-bit & 275  & 0.825  \\
\hline
\multirow{3}{*}{8 kHz} & 8-bit  & 319  & 0.957  \\
                       & 10-bit & 338  & 1.014 \\
                       & 12-bit & 375  & 1.125 \\
\hline
\multirow{3}{*}{16 kHz} 
      & 8-bit   & 489 & 1 467 \\
      & 10-bit  & 518 & 1 554 \\
      & 12-bit  & 575 & 1 725 \\
\hline
\multirow{3}{*}{24 kHz} & 8-bit  & 659  & 1.977 \\
                        & 10-bit & 698  & 2.094 \\
                        & 12-bit & 775  & 2.325 \\
\hline
\end{tabular}
\label{table:adcpowerbit}
\vspace{-01.51800em}
\end{table}

\vspace{-0.85em}
\subsection{Evaluation of power consumption}
\label{subsec:Evaluation for power consumptio}

This work aims to reduce the sampling frequency of hearables so that hearables can save power, which will eventually extend the battery life. However, reducing the sampling frequency reduces the audio quality. Therefore, the mobile platform, such as a smartphone, will use SUBARU to recover high-fidelity audio from the low-resolution signal received from hearables,  restoring audio quality and intelligibility. Sections \ref{subsec:Evaluation for joint BWE and multi-modal SE}, \ref{subsec:inference speed at desktop and Google Pixel7}, \ref{subsec:Evaluation for BWE}, \ref{subsec:Evaluation for only BWE with different sub-Nyquist sampling} evaluate the performance of SUBARU to recover high-fidelity audio in different noisy conditions at different sampling frequencies for real-time streaming operation. However, this section will evaluate how much power savings could be possible if we reduce the sampling frequency and bit resolutions at the ADC of the hearables and how much power will be additionally required to run SUBARU on mobile platforms, such as smartphones. 

Our wearable platform chip NRF52840 has a built-in ADC peripheral, which has a variable sampling frequency up to $\sim$200 ksps and a variable bit resolution up to 12 bits. We vary the sampling frequency from 4 kHz to 24 kHz for 8-bit, 10-bit, and 12-bit resolutions and measure power for each combination. The results are summarized in Table \ref{table:adcpowerbit}. These figures are typical averages under continuous sampling at 3.0 V and assume a single channel with EasyDMA enabled.

To calculate the power consumption in NRF52840's ADC, we first run a workload on NRF52840 that does not have any ADC operations. We call this baseline power. Later, we run the same workload with ADC operations for different sampling frequencies and bit resolutions. Next, we roughly estimate the power consumption by the ADC by subtracting the baseline power from the power found at different sampling frequencies and bit resolutions.  

Table \ref{table:adcpowerbit} indicates that if we use \{4 kHz, 8-bit\} sampling instead of \{24 kHz, 12-bit\} sampling in hearables, we can save 2.325/0.702 = 3.31x power in hearables. That means that we can increase the battery life by $\sim$3.31x for hearables. Therefore, the idea is that SUBARU will reduce the sampling frequency and bit-resolutions in hearables from \{24 kHz, 12-bit\} or \{16 kHz, 12-bit\} to  \{4 kHz, 8-bit\} to save power in hearables. Later, SUBARU will restore the low-resolution audio from \{4 kHz, 8-bit\}  to \{24 kHz, 12-bit\} or \{16 kHz, 12-bit\} on mobile platforms to provide the same audio quality.

\begin{table}[ht!]
\vspace{-01.00em}
    \scriptsize
\setlength{\tabcolsep}{1.3pt}
    \centering
    \caption{Power consumption in Google Pixel7 at different transmission rates to run SUBARU.}
    \vspace{-0.71800em}
    \begin{tabular}{ m{1.5cm}|m{2.3cm}|m{3cm}|l }
    \hline
         \cellcolor [gray]{0.85}\textbf{Transmission rate} & \cellcolor [gray]{0.85}\textbf{Power for SUBARU only} & \cellcolor [gray]{0.85}\textbf{Power for different transmission rates only} &  \cellcolor [gray]{0.85}\textbf{Total power}\\ 
        \hline
        \hline
        64 kbps  & 1.143 W  &  3.54 mW  &  1.146 W \\
        128 kbps  & 1.151 W &   4.18 mW  & 1.155 W\\
        256 kbps  &  1.187 W &   5.98 mW   & 1.193 W\\
        \hline
                  &           &            & Avg. = 1.16 W \\
        \hline
    \end{tabular}
    \vspace{-0.20em}
    \label{table:poeringooglepixel}
    \vspace{-0.85800em}
\end{table}

\textbf{\textit{The next question is how much power is required to restore the audio by SUBARU on mobile platforms.}} We evaluate SUBARU on Google Pixel7 for different transmission rates from hearables to Pixel7 (see Table \ref{table:poeringooglepixel}). The second column in Table \ref{table:poeringooglepixel} indicates the rough estimate of power consumption to run only the SUBARU model on Pixel7, and the 3rd column is the power consumption related only to the reception of packets from the hearables. The power related to only SUBARU increases slightly with the transmission rates because higher transmission rates mean more computation for the time-frequency spectrograms and raw waveforms. The average power consumption of $\sim$1.16 W for running SUBARU on smartphones, such as Pixel7, is trivial compared to the 3.31x increase of power saving in hearables.  Because typically smartphones have a 100x larger battery compared to hearables. \textit{For example,  Samsung Galaxy Buds2 Pro has a 50 mAh battery, and Pixel7 has a 4355 mAh battery. The 1.16 W of power consumption by SUBARU would take around 14 hours to drain the 4355 mAh battery of Pixel7.}

\vspace{-0.50em}
\subsection{Evaluation for only BWE with inference time}
\label{subsec:Evaluation for BWE}

As our SUBARU can do both the BWE and multimodal SE jointly, here, we compare SUBARU with only SOTA BWE models separately for a better understanding of SUBARU's performance at this task. To evaluate the performance for only on the BWE task, we convert SUBARU to only the BWE network by making it a single-input network (i.e., audio from the ACM only). In other words, we feed nothing as a BCM signal to the time enhancement network. We use non-noisy VCTK \cite{yamagishi2019cstr} audio from ACMs to evaluate only the BWE task. Three U-Net-based models, three GAN-based models, one diffusion-based model, and one U-Net + GAN-based model are chosen for comparison. The comparison is shown for desktop and Google Pixel7 in Table \ref{table:comparison_bwe_desktop and cells phone} for 4 kHz to 16 kHz BWE task focusing on the inference time for each model.

\begin{table}[ht!]
\vspace{-0.91800em}
    \scriptsize
\setlength{\tabcolsep}{0.4pt}
    \centering
    \caption{Evaluating BWE for 4 - 16 kHz upsampling for 12-bit resolutions for both desktop and Google Pixel7 platforms. Here, D = Desktop, G = Google Pixel7}
    \vspace{-0.51800em}
    \begin{tabular} { p{01.55cm}| p{01.3cm} | p{01.1cm} |p{01.1cm} |p{01.1cm} |p{01.2cm} |p{01.1cm} }
    \hline
         \cellcolor [gray]{0.85}\textbf{Model} &\cellcolor [gray]{0.85}\textbf{Infer. (ms) (D/G)} & \cellcolor [gray]{0.85}\textbf{L $\downarrow$ (D/G)} & \cellcolor [gray]{0.85}\textbf{V $\uparrow$ (D/G)} & \cellcolor [gray]{0.85}\textbf{N $\uparrow$  (D/G)} & \cellcolor [gray]{0.85}\textbf{S $\uparrow$ (D/G)} & \cellcolor [gray]{0.85}\textbf{P $\uparrow$ (D/G)} \\ 
        \hline
        \hline
        Unprocessed   &   & 2.75/2.79 & 1.92/1.90  &  1.41/1.38 & 9.95/9.89 & 1.23/1.20 \\
         \Xhline{3\arrayrulewidth}
      
       TFiLM \cite{birnbaum2019temporal}  & 4.85/198.87 & 1.65/1.67 &    3.81/3.80   & 3.65/3.63   &  12.58/12.51   &  2.14/2.11   \\
     
        AFiLM\cite{rakotonirina2021self}  & 5.72/235.52 & 1.63/1.66 &   3.82/3.79  &  3.68/3.64      & 12.23/12.17   &  2.12/2.11  \\ 
        
         ATS-UNet \cite{rakotonirina2021self} & 3.79/159.18 &   1.72/1.74    & 3.62/3.59  &  3.41/3.38   &  10.41/10.35    & 1.65/1.64   \\ 
         
          \Xhline{3\arrayrulewidth}
         AERO \cite{mandel2023aero}  &  36/1441.48   &  0.95/0.96   & 4.21/4.19    &   4.12/4.09   &  \textbf{18.15/18.10}  &  2.98/2.96   \\ 
         
         EBEN \cite{hauret2023eben}  &  12.7/520.73  &  1.13/1.16    &   3.93/3.91    & 3.98/3.96    & 16.51/16.23 &  2.85/2.82    \\ 
         
         HiFi++ \cite{kim2023hifi++}  &  6.3/258.31  &  0.85/0.87    &   \textbf{4.26/4.23}   &  \textbf{4.21/4.17}  & 18.13/18.08   &   2.90/2.88    \\ 
         
         NVSR \cite{liu2022neural}   &  54/2268.57 & 0.95/0.98  &  4.11/4.10   &   4.08/4.05   & 17.14/17.11   & 2.64/2.62     \\
         
         NU-Wave \cite{lee2021nu}  &   71/2911.23 & 1.42/1.44  &   3.89/3.85   & 2.35/2.31 &   13.23/13.16  & 2.01/1.99     \\
        \hline
        \hline
       \textbf{SUBARU}  &  \textbf{1.74/70.41} &  \textbf{0.82/0.84} &  \textbf{4.22/4.19}  &   \textbf{4.19/4.13  } &   \textbf{18.11/18.07}  &  \textbf{3.01/2.99}   \\
        \hline
    \end{tabular}
    \vspace{-0.20em}
    \label{table:comparison_bwe_desktop and cells phone}
    \vspace{-0.5800em}
\end{table}

Despite being a U-Net-based model,  SUBARU achieves the lowest LSD compared to all GANs, which is an indication that higher-frequency components are reconstructed in a better way by SUBARU. However, HiFi++  provides slightly better VISQOL and NISQA-MOS, and AERO provides slightly better SI-SDR compared to SUBARU. However, the difference between SUBARU and GANs is minimal for VISQOL, NISQA-MOS, and SI-SDR.  Moreover, SUBARU outperforms all GANs in terms of PESQ and STOI. This is an indication that SUBARU achieves BWE without sacrificing perceptual quality and intelligibility of the audio. 

Please note that every model has a slight performance degradation in Google Pixel7 compared to the desktop due to hardware differences, driver differences between CUDA and ONNX modules, and audio frontend mismatches between the desktop and Pixel7. Despite performance degradation, SUBARU is still best for LSD with the smallest inference time compared to the best-performing GAN-based models. 

\vspace{-0.71800em}
\subsection{Evaluation for only BWE with different sub-Nyquist sampling with music}
\label{subsec:Evaluation for only BWE with different sub-Nyquist sampling}

Section \ref{subsec:Evaluation for BWE} only considers the evaluation of BWE for 4 kHz to 16 kHz upsampling. However, here,  we vary the sampling in different scales, such as 4 kHz to 22 kHz and 8 kHz to 22 kHz, to compare SUBARU's performance on two different datasets: the VCTK \cite{yamagishi2019cstr} (speech data) and MagnaTagATune \cite{law2009evaluation} (music data). The reason for introducing the music dataset is that music has dominant high-frequency components compared to speech, and we want to evaluate SUBARU for high upsampling ratios for both speech and music audio. The detailed comparison is shown in Table \ref{table:BWEwithdifffreqand music}. 

\begin{table}[ht!]
\vspace{-0.51800em}
    \scriptsize
\setlength{\tabcolsep}{1.0pt}
    \centering
    \caption{Evaluating for two different datasets for 12-bit resolutions on the desktop for 4-22 kHz and 8-22 kHz upsampling.} 
    \vspace{-0.51800em}
    \begin{tabular}
     {l|c|c|c|c|c|c|c||c|c|c|c|c|c}
    \hline

         \cellcolor [gray]{0.85}\textbf{Model} & \cellcolor [gray]{0.85}\textbf{Dataset}    & \multicolumn{6}{c||} {\cellcolor [gray]{0.85}\textbf{4 to 22 kHz}} &  \multicolumn{6}{|c} {\cellcolor [gray]{0.85}\textbf{8 to 22 kHz}} \\ 
         \hline
         &   & \textbf{L$\downarrow$} & \textbf{V$\uparrow$}  & \textbf{N$\uparrow$} & \textbf{S$\uparrow$}  & \textbf{P$\uparrow$} & \textbf{ST$\uparrow$} & \textbf{L$\downarrow$} & \textbf{V$\uparrow$}  & \textbf{N$\uparrow$} & \textbf{S$\uparrow$}  & \textbf{P$\uparrow$} & \textbf{ST$\uparrow$}\\
         
        \hline
        Unprocessed  &   &  2.75 & 1.92  &  1.41 & 9.95 & 1.23 &  0.73 &  1.87  &  2.94 & 2.84 & 12.47  &  1.68  & 0.76\\
        \hline

         \multirow{2}{*} {TFiLM \cite{birnbaum2019temporal}} & VCTK   & 1.97   &  3.11    & 2.94    &  10.43    &  1.78 &  0.80  & 1.37   &  4.27    & 4.15    &  16.21    &  2.14 &  0.87  \\ 
          & Magna  & 1.95   &  3.08    & 2.92    &   9.78  &  1.76  &   0.80  &  1.39   &  4.22    & 4.13   &   15.14  &  2.12  &   0.86  \\ 
\hline
          
         \multirow{2}{*} {AERO \cite{mandel2023aero}} & VCTK  &  1.02  &  4.11   &  3.90    &  \textbf{17.49} &  2.77   &   0.88 &  0.89   &  4.39   &  4.21    &  19.14 &  3.15   &   0.91 \\ 
          & Magna & 1.04    &  4.09   &  3.88    &   16.31  &  2.44  &   0.88 &  0.87    &  4.45   &  4.23    &   19.31  &  3.19  &   0.90  \\ 
          
         \hline
        \multirow{2}{*} {EBEN \cite{hauret2023eben}} &  VCTK   &  1.19     &   3.78   & 3.65    & 14.23  & 2.57  & 0.84  &  1.06     &   4.29  & 4.19    & 17.58  & 3.05 & 0.89  \\ 
           &  Magna   & 1.17    &   3.76    & 3.63    & 13.13  & 2.54 & 0.88 &  1.08   &   4.27    &  4.15   & 16.54  & 3.01 & 0.89  \\ 
           
         \hline
        \multirow{2}{*} {HiFi++ \cite{kim2023hifi++}} & VCTK   & 0.91   &   \textbf{4.05}   &  \textbf{4.01}   & 17.29   &  2.84 &  0.89 &   0.80   &   4.44   &  4.26   & 19.11   & 3.13 &  \textbf{0.91} \\ 
          &  Magna  & 0.90   &   \textbf{4.01}   &  \textbf{3.96}   & 16.42   & 2.83   &   0.88 &  0.83   &   4.41   &  4.21   & 18.71   & 3.09   &   \textbf{0.90}   \\ 
          
         \hline
         \multirow{2}{*} {NVSR \cite{liu2022neural}}   &  VCTK  & 1.02  &  3.95    &   3.88   &  16.31  & 2.41 &   0.85  & 0.91   &  4.31    &   4.21   & 17.81    & 3.1 &   0.87  \\
         &  Magna   & 1.04   &  3.91    &   3.85    & 15.88     & 2.40 &  0.85 & 0.93   &  4.28    &   4.17    & 16.48     & 3.0 &  0.86   \\
         
        \hline
        \hline
       \multirow{2}{*} {\textbf{SUBARU}}  &  VCTK  & \textbf{0.86} &  \textbf{4.03}  &   \textbf{3.99} &   \textbf{17.45}    &  \textbf{2.97} & \textbf{0.90}  &  \textbf{0.78} &  \textbf{4.46}  &   \textbf{4.26} &   \textbf{19.75}    &  \textbf{3.33} & \textbf{0.90}  \\
         &  Magna  & \textbf{0.87} &  \textbf{4.01}  &   \textbf{3.97} &   \textbf{16.38}  &  \textbf{2.95}  &  \textbf{0.90} &  \textbf{0.80} &  \textbf{4.44}  &   \textbf{4.25} &   \textbf{18.58}  &  \textbf{3.31}  &  \textbf{0.90} \\
        \hline
    \end{tabular}
    \vspace{-0.20em}
    \label{table:BWEwithdifffreqand music}
    \vspace{-0.700em}
\end{table}

Table \ref{table:BWEwithdifffreqand music} indicates that SUBARU provides comparable BWE for both speech and music datasets with the same inference speed. It indicates that SUBARU can be a generalized solution for both speech and music in hearables. 

Please also note that the metrics improve for 8-22 kHz compared to 4-22 kHz because for 8-22 kHz the upsampling ratio is around 2.75, and for 4-22 kHz the upsampling ratio is 5.5. The lower the upsampling ratio, the greater the performance gain for the reconstruction model. Table \ref{table:BWEwithdifffreqand music} also indicates that our SUBARU outperforms GANs in terms of LSD, PESQ, and STOI, and has very similar performance in VISQOL and NISQA-MOS for both the speech and music datasets.

\vspace{-0.91800em}
\subsection{Different ADC bit resolutions and sampling frequencies}
\label{subsec:Evaluation at different bit resolutions}

Please note again that varying bit resolution means varying  ADC's sampling bit resolution, not changing the model quantization. Here, we vary the bit resolutions and sampling frequencies of the ADC during sampling while doing joint BWE and multimodal SE at noisy conditions and summarize the performance of SUBARU in Table \ref{table:bitresolution and BWE}. We vary the bit resolution from 8 to 12 bits for 4-16 kHz and 4-22 kHz upsampling, and SUBARU is evaluated on the desktop. Table \ref{table:bitresolution and BWE} shows that SUBARU's performance degrades at low bit resolutions because of the increase in the quantization error introduced while sampling. However, SUBARU's performance is still better than that of the best-performing models.

\begin{table}[ht!]
\vspace{-0.81800em}
    \scriptsize
\setlength{\tabcolsep}{0.1pt}
    \centering
    \caption{Evaluating SE for 8, 10, and 12-bit ADC resolutions on the desktop for 4 - 16 kHz and 4 - 22 kHz upsampling with noisy data for the vibration sensor. Here, Param = Parameter and Infer. = Inference time.}
    \vspace{-0.51800em}
    \begin{tabular}
     {m{01.1cm}|m{0.650cm}|m{01cm}|m{01.10cm}|m{01.5cm}|m{01.3cm}|m{01.0cm}|m{01.0cm}}
    \hline

         \cellcolor [gray]{0.85}\textbf{Model} & \cellcolor [gray]{0.85}\textbf{ADC bit}   & \cellcolor [gray]{0.85}\textbf{LSD$\downarrow$ (4-16k/4-22k)} & \cellcolor [gray]{0.85}\textbf{VISQOL \hspace{0.6cm}(4-16k/4-22k)} & \cellcolor [gray]{0.85}\textbf{NISQA-MOS  \hspace{0.0cm}(4-16k/4-22k)} & \cellcolor [gray]{0.85}\textbf{SI-SDR  \hspace{0.6cm}(4-16k/4-22k)} & \cellcolor [gray]{0.85}\textbf{PESQ \hspace{0.6cm} (4-16k/4-22k)} & \cellcolor [gray]{0.85}\textbf{STOI  \hspace{0.6cm} (4-16k/4-22k)} \\ 
         \hline
         
        \hline
        \hline

         \multirow{3}{*} {AERO}  &  8  &  0.99/1.04 &  4.01/3.96   &  3.91/3.80   &  16.03/16.42 &  2.85/2.69  &  0.88/0.87    \\ 
           & 10  & 0.98/1.03    &  4.10/4.06   &  3.98/3.86    &   16.69/17.14  &  2.89/2.73 & 0.89/0.88    \\ 
          & 12  & 0.97/1.02    &  4.16/4.11   &  4.03/3.90   &   17.03/17.49  &  2.93/2.77 & 0.89/0.88    \\ 
       
         \hline
        \multirow{3}{*} {HiFi++}  &  8   & 0.91/93   &   4.01/3.96   &  3.97/3.93   & 16.28/16.14   &  2.81/2.80 &  0.89/0.88  \\ 
         &   10  & 0.90/0.92   &   4.13/3.99   &  4.07/3.97   & 16.95/16.85   & 2.83/2.82   &   0.89/0.88  \\ 
          &   12  & 0.89/0.91   &   4.18/4.05   &  4.11/4.01   & 17.48/17.29   & 2.85/2.84   &   0.90/0.89  \\

        \hline
        \hline
       \multirow{3}{*} {\textbf{SUBARU}}   &  8   & \textbf{0.85/0.87} &  \textbf{4.21/3.91}  &   \textbf{4.05/3.99} &   \textbf{16.84/16.71}    &  \textbf{2.92/2.90} & \textbf{0.90/0.89}  \\
       &   10   & \textbf{0.84/0.86} &  \textbf{4.27/3.99}  &   \textbf{4.11/3.94} &   \textbf{17.11/17.02}  &  \textbf{2.95/2.93}  &  \textbf{0.90/0.90} \\

       &   12   & \textbf{0.84/0.86} &  \textbf{4.35/4.03}  &   \textbf{4.19/3.99} &   \textbf{17.94/17.45}  &  \textbf{3.00/2.97}  &  \textbf{0.90/0.90} \\

        \hline
    \end{tabular}
    \vspace{-0.20em}
    \label{table:bitresolution and BWE}
    \vspace{-0.75800em}
\end{table}

\begin{table}[ht!]
\vspace{-0.951800em}
   \scriptsize
\setlength{\tabcolsep}{0.4pt}
    \centering
    \caption{Evaluation at two different mobile platforms.}
    \vspace{-0.51800em}
    \begin{tabular}
     {c|c|c|c|c|c|c|c}
    \hline

         \cellcolor [gray]{0.85}\textbf{Model} & \cellcolor [gray]{0.85}\textbf{Platform}  & \cellcolor [gray]{0.85}\textbf{ADC resolution} & \cellcolor [gray]{0.85}\textbf{Inference}   &  \cellcolor [gray]{0.85}\textbf{Avg. Power} & \cellcolor [gray]{0.85}\textbf{LSD$\downarrow$} & \cellcolor [gray]{0.85}\textbf{PESQ $\uparrow$} & \cellcolor [gray]{0.85}\textbf{STOI $\uparrow$} \\ 
         \hline
         
        \hline
        \hline

        \multirow{2}{*} {\textbf{SUBARU}}  &   Pixel7   & 12-bit &  70.41 ms  &   1.16 W &     0.85   &  2.97  &  0.90  \\

       &   Galaxy S21   & 12-bit &  99.3 ms  &   1.24 W &   0.85   & 2.97 & 0.90 \\

        \hline
    \end{tabular}
    \vspace{-0.20em}
    \label{table:Samsung vs pixel7}
    \vspace{-0.5800em}
\end{table}

\vspace{-0.70em}
\subsection{Evaluation at other mobile platforms}
\label{subsec:Evaluation at other mobile platforms}

To generalize the performance evaluation of SUBARU, we evaluate SUBARU on Samsung Galaxy S21 and compare its performance with the Google Pixel7. The result is summarized in Table \ref{table:Samsung vs pixel7}. Samsung Galaxy S21 has an Adreno 660 GPU with 8 GB RAM and supports TensorFlow Lite with GPU delegate. Therefore, we can infer SUBARU on the Samsung Galaxy S21 similarly to Google Pixel7. The inference speed on the Samsung Galaxy S21 is 1.41x slower than Google Pixel7, since Pixel7 has a custom tensor processing unit (TPU) in its Tensor G2 chipset. The evaluation metrics are similar to Pixel7, as both the Pixel7 and Galaxy S21 use the same SUBARU model with the same quantization (float 32). The power is slightly lower for Pixel7 as its TPU processing requires less power compared to the GPU in Galaxy S21.

\vspace{-01.3100em}
\subsection{Ablation study on the model components}
\label{subsec:Ablation study on the model}

In this section, we conduct an ablation study on several important modules in SUBARU to understand the contribution of each module to overall performance. We evaluate the impact of the important modules using the VCTK dataset in noisy conditions, with the results summarized in Table \ref{table:Ablation study on the model}. Table \ref{table:Ablation study on the model} shows that replacing Mamba in our spectral enhancement network and time enhancement network yields similar performance metrics (LSD, SI-SDR, PESQ, STOI). However, the training time increases significantly from 232 s to 389 s per epoch, and the number of parameters increases slightly from 3.61 M to 3.74 M. Therefore,  we keep Mamba in our design as Mamba provides similar performance with a shorter training time and smaller parameters.

\begin{table}[ht!]
\vspace{-0.821800em}
    \scriptsize
\setlength{\tabcolsep}{1.3pt}
    \centering
    \caption{Ablation study for model components for 4 - 16 kHz upsampling on the desktop platform for 12-bit resolutions on the VCTK dataset in noisy conditions for the vibration sensor.}
    \vspace{-0.71800em}
    \begin{tabular}{ m{3cm}|l|l|l|l| m{1.4cm} | m{1.4cm} }
    \hline
         \cellcolor [gray]{0.85}\textbf{Method} & \cellcolor [gray]{0.85}\textbf{L $\downarrow$} & \cellcolor [gray]{0.85}\textbf{S $\uparrow$} & \cellcolor [gray]{0.85}\textbf{P $\uparrow$} & \cellcolor [gray]{0.85}\textbf{ST $\uparrow$} & \cellcolor [gray]{0.85}\textbf{Parameter} & \cellcolor [gray]{0.85}\textbf{Train time per epoch}\\ 
        \hline
        \hline
        SUBARU  & 0.84  & 17.94  & 3.00  & 0.90 & 3.61 M  &  232 s     \\
        \hline
        Replace Mamba with Transformers in the spectral and time enhancement networks  & 0.82  & 18.12  & 3.07  & 0.90  & 3.74 M  &  389 s  \\
        \hline
        without time enhancement network  &  0.86  & 15.87  & 2.98  & 0.90 &  2.8495 M & 189 s  \\
        \hline
        without amplitude-phase enhancement network  &  1.28  &  8.12 &  2.05  & 0.81 & 2.7814 M &  155 s\\
        \hline
        Changing the order of time and amplitude-phase enhancement network  & 0.84  & 17.94  & 3.00  & 0.90 & 3.61 M  &  232 s   \\
        \hline
    \end{tabular}
    \vspace{-0.20em}
    \label{table:Ablation study on the model}
    \vspace{-0.75800em}
\end{table}

We can see from Table \ref{table:Ablation study on the model} that the time enhancement and amplitude-phase enhancement networks are two important components, as their absence notably degrades the model performance for all metrics, and they have an impact on both the BWE and SE tasks. The reason behind this is that the time enhancement network improves the low-resolution, noisy data in the time domain, and the amplitude-phase enhancement network reconstructs clean phases in noisy conditions. 
Moreover, as the time enhancement network and the amplitude-phase enhancement network can take a raw waveform as input and give a raw waveform as output, we can change the order of these two networks. However, changing the order of these two networks does not change the model's performance.

\begin{table}[ht!]
\vspace{-0.51800em}
    \scriptsize
\setlength{\tabcolsep}{1.1pt}
    \centering
    \caption{Ablation study for loss functions for 4 - 16 kHz upsampling on the desktop platform for 12-bit resolutions on the VCTK dataset in the noisy condition.}
    \vspace{-0.51800em}
    \begin{tabular}{ m{5cm}|l|l|l|l }
    \hline
         \cellcolor [gray]{0.85}\textbf{Method} & \cellcolor [gray]{0.85}\textbf{LSD $\downarrow$} & \cellcolor [gray]{0.85}\textbf{SI-SDR $\uparrow$} & \cellcolor [gray]{0.85}\textbf{PESQ $\uparrow$} & \cellcolor [gray]{0.85}\textbf{STOI $\uparrow$}\\ 
        \hline
        \hline
        SUBARU  &   0.84  & 17.94  & 3.00  & 0.90  \\
        Without multi-resolution STFT loss function   &  1.75  & 8.57  & 1.85  & 0.82  \\
        with only multi-resolution STFT loss  & 0.93  & 14.25 & 2.43 & 0.86 \\
        with only multi-resolution STFT + multi-scale loss  & 0.87  & 15.28 & 2.74 & 0.87 \\
        with only multi-resolution STFT + multi-scale + multi-period loss  & 0.85  & 16.75 & 2.89 & 0.88 \\
        with multi-resolution STFT + multi-scale + multi-period loss + phase spectrum loss & 0.84  & 17.94  & 3.00  & 0.90 \\
         
       
        \hline
    \end{tabular}
    \vspace{-0.20em}
    \label{table:Ablation study on the loss}
    \vspace{-01.85800em}
\end{table}

\vspace{-0.5800em}
\subsection{Ablation study on the loss function}
\label{subsec:Ablation study on the loss function}

We evaluate the impact of the loss functions using the VCTK dataset in noisy conditions, with the results summarized in Table \ref{table:Ablation study on the loss}. If the multi-resolution STFT loss function is removed, the model performance degrades drastically. The reason behind this is that the multi-resolution STFT loss function is the only loss function in the frequency domain in SUBARU, and without it, the relationship among the high- and low-frequency components cannot be learned effectively in noisy conditions between the low-resolution and high-resolution spectral frames.  If multi-scale and multi-period loss functions are added one at a time with the multi-resolution STFT loss, the performance increases gradually, similar to multi-scale and multi-point discriminators in GANs. Multi-scale and multi-period loss functions work in the time domain and facilitate learning the time-domain low-resolution to high-resolution mapping effectively. 

\vspace{-01.281800em}
\subsection{Evaluation on live noise data}
\label{subsec:Evaluation on live noise data}

In earlier sections, we simulated noisy speech by adding noise directly to clean recordings. In contrast, this section focuses on evaluating SUBARU in real-world conditions, where background noise occurs naturally and continuously. Since clean reference signals are unavailable in such live settings, we calculate the Character Error Rate (CER) using Whisper \cite{radford2022whisper} to assess performance. We conduct tests using five volunteers per scenario, each reading 20 consistent sentences. This consistency across tests allows us to analyze the effect of various environmental conditions on SUBARU's accuracy. We test SUBARU inside and outside the lab in different scenarios.

\subsubsection{Inside the lab:}  We design the following four scenarios to evaluate SUBARU inside the lab:

\begin{itemize}
    
\item \textbf{Quiet Room:} To provide a baseline for comparison, volunteers speak in a quiet environment.

\item \textbf{Live speech:} While one volunteer speaks, another volunteer stands one meter away and talks simultaneously, simulating a live speech interference. The sound pressure level (SPL) for the target and interfering speakers are approximately 77 dB and 54 dB, respectively.

\item \textbf{Music:} During the speech, two speakers on either side of the participant play music with SPLs of 65 dB and 64 dB, respectively. This setup introduces multimedia noise to test the system’s robustness.

\item  \textbf{Mobile scenarios:} Volunteers speak while engaging in physical activity, such as running on a treadmill, cycling on a stationary bike, or walking. 
\end{itemize}

\subsubsection{Outside the lab:}  We design the following three scenarios to evaluate SUBARU outside the lab:

\begin{itemize}
    
\item \textbf{Bus ride:} The target volunteer reads while sitting on a bus with typical background road-noise from the surrounding environment. The average SPL is 44 dB.

\item \textbf{Classroom:} The target volunteer reads while surrounded by people talking in a crowded classroom. The average SPL is 63 dB.

\item \textbf{Car ride:} The target volunteer reads in a car with an open window. The average SPL is 61 dB.

\end{itemize}

Table \ref{table:livenoise} summarizes the results. In all noisy scenarios, the unprocessed speech signal yields a high CER, particularly in the presence of competing speech and a car ride with an open window. After introducing SUBARU, the enhanced signals show a marked improvement in recognition accuracy, indicating effective suppression of live acoustic interference. Additionally, in mobile conditions, user movement has some impacts on the performance because of the presence of vibrations related to movements. To prevent this, we use a high-pass filter (HPF) with a cut-off frequency of 15 Hz to remove the movement-induced noises from the BCM. The results demonstrate that SUBARU can be used in voice communication in noisy public settings.

\begin{table}[ht!]
\vspace{-0.51800em}
   \scriptsize
\setlength{\tabcolsep}{0.9pt}
    \centering
    \caption{The CER with live noises for 4 - 16 kHz upsampling for the vibration sensor only.}
    \vspace{-0.51800em}
    \begin{tabular}
     {c|m{0.7cm}|m{0.8cm}|c|m{2cm}|m{0.8cm}|c|c|}
    \hline

              & \multicolumn{4}{c|} {\cellcolor [gray]{0.85}\textbf{Inside the lab}} &  \multicolumn{3}{c} {\cellcolor [gray]{0.85}\textbf{Outside the lab}} \\ 
         \hline
            &  \textbf{Quiet room}  &  \textbf{Live speech} & \textbf{Music}  & \textbf{Mobile Scenarios} & \textbf{Bus ride}  & \textbf{Classroom} & \textbf{Car ride}\\
         
        \hline
        \hline
         Raw speech  &  0.05 & 0.74   & 0.41  & 0.38 (before HPF) / 0.24 (after HPF) & 0.61 & 0.67 & 0.79  \\

     \hline

        SUBARU  &  0.05 & 0.33   & 0.25  & 0.09 (before HPF) / 0.07 (after HPF) & 0.28 & 0.29 & 0.37  \\
        \hline
    \end{tabular}
    \vspace{-0.20em}
    \label{table:livenoise}
    \vspace{-0.5800em}
\end{table}

\vspace{-01.281800em}
\subsection{Subjective Analysis}
\label{subsec:Subjective Analysis}

\begin{figure}[ht!]
  \centering
 \includegraphics[width=0.3\textwidth,height=0.12\textheight]{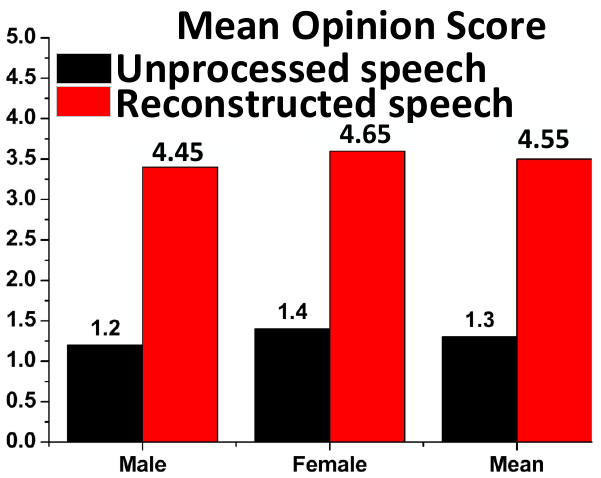} 
 \vspace{-01.05em}
  \caption{Results of MOS. }
  \label{fig:comparison}
  \vspace{-0.9515em}
\end{figure}

We conducted a listening study to assess perceived quality across 12-24 kHz extension range for  8 bit resolution using the 5-point Mean Opinion Score (MOS) ratings. A panel of 12 trained listeners evaluated each condition—Unprocessed, our Proposed method SUBARU, and the clean reference—presented in randomized order. For MOS, listeners rated each sample on a scale from 1 (bad) to 5 (excellent). As shown in Fig. \ref{fig:comparison}, the unprocessed signals received the lowest MOS across all ranges (mean scores around 1.3), reflecting severe information loss. Our Proposed method achieved substantial improvements—mean MOS of approximately 4.55 narrowing the gap to the clean condition.

\vspace{-0.9705em}

\section{Limitation}
\label{sec:discussion}

\subsection{Encryption and Codec}

This work does not consider any encryption of the sampled audio before transmission to the mobile platform from the hearables. Moreover, we do not include any codec while evaluating its performance, power, and efficiency. Our future work will include codec and encryption while designing a joint BWE and multimodal SE algorithm. However, as a codec is a compression and decompression algorithm, codec can be used safely after sub-Nyquist sampling on hearables with our platform.


\vspace{-1.2em}
\subsection{Fine tuning on mobile platform}

We fine-tune SUBARU on a desktop with GPU support due to the limitation of available open-source tools and computational resources for fine-tuning models on a mobile platform. Our future work will try to solve this problem by contributing to open-source tool development for on-device training.

\vspace{-1.2em}
\section{Conclusion}
\label{sec:Conclusion}

SUBARU is designed to be deployed in a split architecture, where hearables run sub-Nyquist sampling in low bit resolution on signals from both ACMs and BCMs, and mobile platforms connected to hearables run the joint BWE and multimodal SE algorithms to reconstruct high-resolution, noise-free audio in noisy conditions. The low latency of SUBARU while performing joint BWE and multimodal SE  enables streaming enhancement, low power, and low memory solutions, making SUBARU deployable on mobile platforms. 
Moreover, SUBARU adopts a hybrid architecture by merging both waveform-based and spectrum-based methods, enabling joint training in both spectrum and waveform domains. This joint training also improves both time and spectral domain features during audio reconstruction. This joint training also enables the use of instantaneous phase and group delay anti-wrapping losses to reconstruct clean phases in noisy conditions.  SUBARU provides users with reliable communication while ensuring real-time performance by extending the battery life of hearables. Therefore, SUBARU is designed to meet the high demands of smart hearables in low-power and real-world usage by bridging the gap between power and the performance of current commercial technologies.

\vspace{-0.500em}
\section{Acknowledgment}
\vspace{-0.3em}

The research is supported by the Grant no: N000142612163 from the Office of Naval Research (ONR).

\bibliographystyle{IEEEtran}
\bibliography{IEEE_TASLP/main_v1}

\end{document}